\documentclass[]{iopart}
\usepackage{iopams}
\usepackage{dsfont}

\newcommand{\dv}[1]{\partial_{#1}}

\newcommand{\demi}{\frac{1}{2}}

\newcommand{\dbi}{\partial_{i}}
\newcommand{\dbj}{\partial_{j}}
\newcommand{\dhi}{\partial^{i}}
\newcommand{\dhj}{\partial^{j}}
\newcommand{\dbk}{\partial_{k}}

\newcommand{\dhk}{\partial^{k}}

\newcommand{\mf}[1]{\mathcal{#1}}
\newcommand{\dsd}[2]{\frac{\partial {#1}}{\partial {#2}}}
\newcommand{\eb}{\bar{e}}
\newcommand{\Lie}{\mathcal{L}}
\def\dd{{\rm d}}

\def\HH{\mathcal{H}}
\def\Tr{\mathcal{T}}

\newcommand{\ChristoffelOrdre}[4]{{}^{#4}\Gamma^{#1}_{  #2 #3}}

\begin{document}

\title{Gauge-invariant Boltzmann equation and the fluid limit}

\author{Cyril Pitrou}

\address{Institut d'Astrophysique de Paris, 
Universit\'e Pierre~\&~Marie Curie - Paris VI,
CNRS-UMR 7095, 98 bis, Bd Arago, 75014 Paris, France.}
\ead{pitrou@iap.fr}

\date{today}
\begin{abstract}
This article investigates the collisionless Boltzmann equation up to second order in the
cosmological perturbations. It describes the gauge dependence of the
distribution function and the construction of a gauge-invariant distribution
function and brightness, and then derives the gauge-invariant fluid limit.  
\end{abstract}
 \maketitle
\section{Introduction}\label{sec-intro}

The origin of the large-scale structure is nowadays understood from the gravitational
collapse of initial density perturbations which were produced by amplification
of the quantum fluctuations in the inflaton field \cite{Mukhanov1992}. The properties of the large-scale
structure depend both on the initial conditions at the end of inflation and on the growth of perturbations in
a universe filled with non relativistic matter and radiation. The theory of cosmological perturbations is thus a cornerstone of our
understanding of the large-scale structure. The evolution of radiation
(photons and neutrinos) needs to be described by a Boltzmann equation \cite{Ehlers-71,Stewart71,Bernstein}. Two
types of perturbative schemes have extensively been used in the literature in
order to describe the evolution of the cosmological perturbations. The first is a $1+3$ covariant splitting of space-time \cite{Challinor_Lasenby,Challinor,elst} and the second is a
more pedestrian coordinate based approach. In the first approach, exact equations on
the physical space-time are derived and perturbative solutions around a
background solution are then calculated. In the second approach, an averaging procedure
is implicitly assumed and, starting from a background space-time, perturbation variables
satisfying the equations of motion order by order are constructed. In the $1+3$ approach, the variables used
are readily covariant, but the absence of background space-time can be a
problem to simplify the resolution by performing a mode expansion,
since the Helmholtz function is in general not defined on the physical space-time. In the
coordinate based approach, all perturbation variables live on the background
space-time, and enjoy the advantages of its highly symmetric properties.
However, this extra mathematical structure is at the origin of the gauge issue
through the identification mapping that we needs to be defined between the
background space-time and the physical space-time. Thus, the gauge dependence
needs to be understood. An elegant solution is to construct gauge-invariant variables {\it
  \`a la Bardeen} both for the metric perturbation variables \cite{bardeen81} and
for the distribution function \cite{Ruth,Ruth2}. Since the Boltzmann and
Einstein equations are covariant, they can be expressed solely in terms of gauge invariant
variables provided we have a full set at hand. A full comparison of these two
formalisms has been performed at first order in Ref.~\cite{Dunsby_comparison}, and for gravitational waves at second order in Ref.~\cite{Bob}.

In the coordinate based approach, the true degrees of freedom identified from the Lagrangian formalism, are
quantized. They transfer to classical perturbations which inherit a nearly
scale invariant power spectrum and Gaussian statistics, when their wavelength
stretches outside the horizon, thus providing initial conditions for the
standard big-bang model. Conserved quantities \cite{MalikWands,Rigopoulos}
enable to ignore the details of the transition between inflation and the
standard big-bang model (see however \cite{BKU2003}), and the evolution
details need only to be known when the wavelength reenters the horizon. A
first step to extend this procedure in the $1+3$ formalism has been taken in
Ref.~\cite{LangloisVernizzi} where conserved quantities were defined. As for
the degrees of freedom which need to be quantized, a first proposal was made
in Ref.~\cite{Pitrou2007}, in order to identify them, but it has not yet been motivated by a Lagrangian formulation. 

The properties of the observed cosmic microwave background (CMB) anisotropies have confirmed the
validity of the linear perturbation theory around a spatially homogeneous and isotropic
universe and have set strong constraints on the origin of structures, as
predicted by inflation. It now becomes necessary, with the forthcoming increasing precision of data that may allow to detect deviation from Gaussianity \cite{SpergelWMAP}, to study the second-order approximation, in order to discuss the accuracy of
these first-order results. These non-Gaussian features are also of first
importance, since they can help discriminating between different inflation
theories. Indeed, one-field driven inflation leads to very small levels of
primordial non-Gaussianity \cite{maldacena03}, whereas multifield inflation can present
significant non-Gaussian features \cite{Bartolo:2001},\cite{BU}. However, since non-Gaussian effects
also appear through non-linear evolution, that is from the second-order
approximation and beyond of the evolution equations, the study of second-order
evolution equations is necessary in order to distinguish between
primordial and evolutionary non-Gaussianities (see Ref.~\cite{BartoloNG} for a
review on non-Gaussianity). Second-order Einstein and Boltzmann
equations have been written in the $1+3$ formalism \cite{Ellis_Boltzmann,Maartens}, but
not integrated numerically, partly because the mode expansion is not defined
on the physical space-time, and this would then require a four dimensional
numerical integration. However, the promising formalism of
Ref.~\cite{Vernizzi_etal}, which builds a bridge between the $1+3$ formalism
and the coordinate based approach, might shed some light on these issues.
Similarly, in the coordinate based approach, the second-order Einstein
equations have been written in terms of gauge-invariant variables
\cite{Nakamura-2006}, and a first attempt has been made to write the Boltzmann
equation in a given gauge for the different species filling the universe, and to solve them analytically \cite{BartoloBoltzmann2_I,BartoloBoltzmann2_II}.

The goal of this paper is to provide the full mathematical framework for
handling distribution functions at second order in the coordinate based
approach taking into account the gauge issue. This will clarify the existing
literature and point out some existing mistakes. We first review briefly in
section II the gauge transformations and the procedure to build gauge
invariant variables. We then present in section III the transformation
properties of the distribution function, and express them up to second order.
We define in section IV the gauge-invariant distribution function and the gauge
invariant brightness up to second order in the particular case of radiation
(but this is readily extendable to cold dark matter). We then deduce in
section V, from the Boltzmann equation, the evolution of the gauge invariant
brightness in its simplest collisionless form, at first and second orders. To finish, we express in
section VI the fluid limit as a consistency check of our results.


\section{Overview on gauge transformations and gauge-invariant variables}\label{sec1}

\subsection{First- and second-order perturbations}\label{sec1part1}

We assume that, at lowest order, the universe is well described by a
Friedmann-Lema\^{\i}tre space-time (FL) with flat spatial sections. The most
general form of the metric for an almost FL universe is

\begin{eqnarray}\label{metric}
 \dd s^2 &=& g_{\mu\nu}\dd x^\mu\dd x^\nu \\
&=& a(\eta)^2 \big\{-(1 + 2\Phi )\dd\eta^2 + 2
 \omega_{i} \dd x^{i}\dd\eta + [(1-2 \Psi)\delta_{ij} + h_{ij}]\dd x^{i}\dd x^{j}\big\},\nonumber
\end{eqnarray}
where $\eta$ is the conformal time and $a$ the scale factor. We perform a
scalar-vector-tensor decomposition as
\begin{equation}\label{eq:I-2}
 \omega_{i}=\dbi B + B_{i}\,,
\end{equation}
\begin{equation}
 h_{ij}=2 E_{ij} + \dbi E_j + \dbj E_i + 2 \dbi \dbj E,
\end{equation}\label{eq:I-3}
where $B_i$, $E_i$ and $E_{ij}$ are transverse ($\dhi E_i= \dhi B_i=\dhi E_{ij}=0$), and $E_{ij}$
is traceless ($E^i_{\,\,i}=0$). There are four scalar degrees of freedom
($\Phi,\,\Psi,\,B,\,E$), four vector degrees of freedom ($B_i,\,E_i$) and two
tensor degrees of freedom ($E_{ij}$). 
Each of these perturbation variables can be split in first and second-order parts as
\begin{equation}\label{decomposition-ordre2}
 W=W^{(1)}+\frac{1}{2}W^{(2)}\,.
\end{equation}
This expansion scheme will refer, as we shall see, to the way gauge transformations and gauge-invariant (GI) variables are defined. First-order variables are solutions of first-order
equations which have been extensively studied (see Ref.~\cite{Uzan-Peter} for a
review). Second-order equations will involve purely second-order terms, e.g.
$W^{(2)}$ and terms quadratic in the first-order variables, e.g.
$[W^{(1)}]^2$. There will thus never be any ambiguity about the order of perturbation variables involved as long as the
order of the equation considered is known. Consequently, we will often omit to specify the order
superscript when there is no risk of confusion.

At first order, 4 of the 10 metric perturbations are gauge degrees of freedom and
the 6 remaining degrees of freedom reduce to 2 scalars, 2 vectors and 2 tensors. The three types of
perturbations decouple and can thus be treated separately. As long as no vector
source terms are present, which is generally the case when no magnetic
field or topological defect is taken into account, the vector modes decay as
$a^{-2}$. Thus, we can safely discard them and set $E^{(1)}_i=B^{(1)}_i=0$. In
the following of this work, we shall not include vector modes for the sake of
clarity. We checked that our arguments and derivation can trivially (but
at the expense of much lengthy expressions) take them into account. 

In the fluid description, we assume that the matter content of the universe can be described by a
mixture of fluids. The four-velocity of each fluid is decomposed as
\begin{equation}
 u^{\mu}=\frac{1}{a}(\delta_{0}^{\mu} + v^{\mu}).
\label{defvelocity}
\end{equation}
The perturbation $v^{\mu}$ has only three independent degrees of freedom since 
$u^{\mu}$ must satisfy $u_{\mu}u^{\mu}=-1$. The spatial components can be decomposed as
\begin{equation}\label{decv}
 v^{i}=\partial^{i}v + \bar{v}^{i}\,,
\end{equation}
$\bar{v}^{i}$ being the vector degree of freedom, and $v$ the scalar degree of
freedom. The stress-energy tensor of this fluid is of the form 
\begin{equation}
 T_{\mu\nu} = \rho u_\mu u_\nu + P\left( g_{\mu\nu} + u_\mu u_\nu\right)\,,
\label{defstressenergytensor}
\end{equation}
where the density and pressure are expanded as follows
\begin{equation}
 \rho = \bar\rho + \delta\rho,\qquad
 P=\bar P + \delta P.
\label{defdeltamatter}
\end{equation}
At the background level, the form of the stress-energy tensor is completely fixed by the symmetry properties
of the FL space-time. However, at the perturbation level, one must consider an anisotropic stress component,
$\pi_{\mu\nu}$ with $\pi_{\,\,\mu}^\mu=u^\mu\pi_{\mu\nu}=0$. The pressure and density of the fluid are related by an equation of state, $P=\rho/3$, in the case of radiation.

At first order, the formalism developed by the seminal work of Ref.~\cite{bardeen81} provides a full set of gauge-invariant variables (GIV). Thanks to
the general covariance of the equations at hand (Einstein equations,
conservation equations, Boltzmann equation), it was shown that it was possible to get first-order equations involving only these
gauge-invariant variables. In addition, if these gauge invariant
variables reduce, in a particular gauge, to the perturbation variables that we
use in this particular gauge, then the computation of the equation can be
simplified. Actually, we only need to compute the equations in this particular
gauge, as long as it is completely fixed, and then to promote by identification our perturbation variables to the
gauge-invariant variables. Thus, provided we know this full set of gauge
invariant variables, the apparent loss of generality by fixing the
gauge in a calculation, is in fact just a way to simplify computations.
Eventually we will reinterpret the equations as being satisfied by gauge
invariant variables. The full set of first-order gauge-invariant variables is
well known and is reviewed in Ref.~\cite{Uzan-Peter} and Ref.~\cite{KodamaSasaki}. As gauge
transformations up to any order were developed, it remained uncertain
\cite{Bruni-1997}, whether or not a full set of gauge-invariant variables
could be built for second and higher orders. This has been recently clarified
\cite{Nakamura-2006}, and the autosimilarity of the transformation rules for
different orders can be used as a guide to build the gauge-invariant variables
at any order. We present a summary of the ideas presented in Ref.~\cite{Bruni-1997} about gauge transformations and the construction of gauge-invariant variables \cite{Nakamura-2006}.

\subsection{Points identification on manifolds}

When working with perturbations, we consider two manifolds: a background
manifold, $\mf{M}_0$, with associated metric $\bar{g}$, which in our case is
the FL space-time, and the physical space-time $\mf{M}_1$ with the metric $g$.
Considering the variation of metric boils down to a comparison between tensor
fields on distinct manifolds. Thus, in order to give a sense to ``$\delta g
(P)= g(P) - \bar{g}(\bar{P})$'', we need to identify the points $P$ and $\bar{P}$ between these two
manifolds and also to set up a procedure for comparing tensors. This will also
be necessary for the comparison of any tensor field.

One solution to this problem \cite{Bruni-1997} is to consider an embedding
$4+1$ dimensional manifold $\mf{N} = \mf{M} \times [0,1]$, endowed with the
trivial differential structure induced, and the projections
$\mathcal{P_{\lambda}}$ on submanifolds  with $\mathcal{P}_0(\mf{N})=\mf{M}
\times \{0\}={\cal M}_0$ and $\mathcal{P}_1(\mf{N})=\mf{M} \times \{1\}={\cal M}_1$. The collection of ${\cal M}_{\lambda}\equiv{\cal
  P}_{\lambda}({\cal N})$ is a foliation of ${\cal N}$, and each element is
diffeomorphic to the physical space-time ${\cal M}_1$ and the background space-time ${\cal M}_0$. The gauge choice on this stack of space-times is defined as a vector field $X$ on $\mf{N}$ which satisfies $X^4=1$ (the component along the space-time slicing $\mathds{R}$). 
A vector field defines integral curves that are always tangent to the vector field
itself, hence inducing a one parameter group of diffeomorphisms
$\phi(\lambda,.)$, also noted $\phi_{\lambda}(.)$, a flow, leading in our case from $\phi(0,p \in
\mathcal{P}_0(\mf{N}))= p \in \mathcal{P}_0(\mf{N})$ along the integral curves
to  $\phi\left(1,p \in \mathcal{P}_0(\mf{N})\right)= q \in
\mathcal{P}_1(\mf{N})$. Due to the never vanishing last component of $X$,
the integral curves will always be transverse to the stack of space-times and
the points lying on the same integral curve, belonging to distinct
space-times, will be identified. Additionally  the property $X^4=1$ ensures
that $\phi_{\lambda,X}(\mathcal{P}_0(\mf{N}))= \mathcal{P}_{\lambda}(\mf{N})$,
i.e. the flow carries a space-time slice to another. This points
identification is necessary when comparing tensors, but we already see that
the arbitrariness in the choice of a gauge vector field $X$ should not have
physical meaning, and this is the well known {\it gauge freedom}.   

\subsection{Tensors comparison and perturbations}

The induced transport, along the flow, of tensors living on the tangent
bundle, is determined by the push-forward $\phi_{\star \lambda}$ and the
pull-back $\phi^{\star}_{\lambda}$ \cite{Wald} associated with an element
$\phi_{\lambda}$ of the group of diffeomorphisms. These two functions encapsulate the transformation properties of the tangent and co-tangent spaces at each point and its image. Indeed, the pull-back can be linked to the local differential properties of the vector field embedded by the Lie derivatives along the vector field in a Taylor-like fashion (see Ref.~\cite{Wald} or Ref.~\cite{Bruni-1997})
\begin{equation}
\Phi^{\star}_{X,\lambda}(T)= \sum_{k=0}^{k=\infty} \frac{\lambda^k}{k!}\Lie_X^k T,
\label{expansion-pullback}
\end{equation}  
for any tensor $T$.

A remark about coordinates changes is on order here. When the tensor  $T$ is a coordinate $x^{\mu}$ (once $\mu$ is fixed, it is a scalar field), the previous definition reduces to the standard finite coordinates transformation.
\begin{equation}
  x'^{\mu} \equiv \Phi^{\star}_{\lambda,\xi}(x^{\mu})= x^{\mu} +\lambda
  \xi^{\mu} + \frac{\lambda^2}{2}\xi^{\mu}_{ ,\nu} \xi^{\nu}+ \dots
\end{equation}
This is the standard way of defining an active transformation on the manifold, by transporting a point of coordinates $x^{\mu}$ to a point of coordinates $x'^{\mu}$. This transformation, when performed on the coordinate system - considering the coordinates as a grid on the manifold that one would displace according to the active transformation - induces a passive coordinates transformation, if we decide that the new coordinates of a point $q$ are the coordinates of the point $p$ such that $\phi_{\lambda}(p)=q$. When considering a transformation induced by a field $\xi$, we will refer to the passive coordinates transformation induced by the active transportation of the coordinates system.

The expansion of Eq.~(\ref{expansion-pullback}) on $\mathcal{P}_0(\mf{N})$ provides a way to compare a tensor field on $\mathcal{P}_{\lambda}(\mf{N})$ to the corresponding one on the background space-time $\mathcal{P}_{0}(\mf{N})$. The background value being $T_0 \equiv \Lie_X^0 T|_{\mathcal{P}_0(\mf{N})}$, we obtain a natural definition for the tensor perturbation 
\begin{equation}
\Delta_X T_\lambda \equiv \sum_{k=1}^{k=\infty} \frac{\lambda^k}{k!}\Lie_X^k T  \Big|_{\mathcal{P}_0(\mf{N})}= \Phi^{\star}_{X,\lambda}(T) - T_0.
\label{perturbation-tenseur}
\end{equation} 
The subscript $X$ reminds the gauge dependence. We can read the $n$-th order
perturbation as
\begin{equation}
\delta^{(n)}_XT \equiv \Lie_X^n T \Big|_{\mathcal{P}_0(\mf{N})}\,,
\end{equation}
which is consistent with the expansion of perturbation variables of the physical metric in Eq.~(\ref{decomposition-ordre2}), since the physical space-time is labeled by $\lambda=1$. However, the fact that the intermediate space-time slices ${\cal P}_{\lambda}({\cal N})$ are labeled by $\lambda$ removes the absolute meaning of order by order perturbations, as it can be seen from Eq.~(\ref{perturbation-tenseur}). The entire structure embedded by $\mf{N}$ is more than just a convenient construction and this will have important consequences in gauge changes as we will now detail.

\subsection{Gauge transformations and gauge invariance}

If we consider two gauge choices $X$ and $Y$, a gauge transformation from $X$ to $Y$ is defined as the diffeomorphism 
\begin{equation}
\phi_{X \rightarrow Y, \lambda} = (\phi_{X,\lambda})^{-1}(\phi_{Y,\lambda}),
\end{equation}
and it induces a pull-back which carries the tensor $\Delta_X T_{\lambda}$, which is the perturbation in the gauge $X$, to $\Delta_Y T_{\lambda}$, which is the perturbation in gauge $Y$ since

\begin{eqnarray}
 \phi_{X \rightarrow Y, \lambda}^{\star}  \left(\Delta_X T_{\lambda}+T_0 \right)&=&\left[(\phi_{X,\lambda})^{-1}(\phi_{Y,\lambda})\right]^{\star}\phi^{\star}_{X,\lambda}(T)\nonumber\\
&=&\phi^{\star}_{Y,\lambda}(\phi^{\star}_{X,\lambda})^{-1}\phi^{\star}_{X,\lambda}(T)\nonumber\\
&=&\phi^{\star}_{Y,\lambda}(T) \nonumber\\
&=&\Delta_Y T_{\lambda}+T_0.
\end{eqnarray}
As demonstrated in Ref.~\cite{Bruni-1997} this family (indexed by $\lambda$)
of gauge transformations fails to be a one parameter group due to the lack of the composition rule. It should be Taylor expanded using the so called knight-diffeormorphism along a sequence of vector fields $\xi_i$. For the three first orders, the expression of this knight-diffeomorphism is 

\begin{eqnarray}
\Phi^{\star}_{Y,\lambda}(T) &=& \phi^{\star}_{X \rightarrow Y, \lambda} \Phi^{\star}_{X,\lambda}(T)  \\\label{knight}
&=&   \Phi^{\star}_{X,\lambda}(T) + \lambda \Lie_{\xi_1}  \Phi^{\star}_{X,\lambda}(T)  + \frac{\lambda^2}{2!} (\Lie_{\xi_2} +\Lie_{\xi_1}^2) \Phi^{\star}_{X,\lambda}(T) \nonumber\\
&& \qquad + \frac{\lambda^3}{6} (\Lie_{\xi_3} + 3 \Lie_{\xi_1}\Lie_{\xi_2}+ \Lie_{\xi_1}^3) \Phi^{\star}_{X,\lambda}(T) .\nonumber
\end{eqnarray}
The vector fields $\xi_1$, $\xi_2$ and $\xi_3$ are related to the gauge vector fields $X$ and $Y$ by $\xi_1= Y-X$, $\xi_2=[X,Y]$ and $\xi_3=[2 X -Y,[X,Y]]$.
By substitution of the perturbation by its expression in Eq.~(\ref{perturbation-tenseur}), we identify order by order in $\lambda$, and obtain the transformation rules for perturbations order by order. The first and second order transformation rules, on which we will focus our attention, are

\begin{eqnarray}
\delta^{(1)}_Y T -\delta^{(1)}_X T &=& \Lie_{\xi_1} T_0, \nonumber\\
\delta^{(2)}_Y T -\delta^{(2)}_X T &=& 2 \Lie_{\xi_1} \delta^{(1)}_X T + ( \Lie_{\xi_2}+ \Lie_{\xi_1}^2)T_0.
\label{transforule}
\end{eqnarray}

The fact that we had to follow $n$ integral curves of $n$ distinct vector fields for $n$-th order perturbations is a characteristic of knight-diffeomorphisms. It arises from the fact that, for the whole differential structure of $\mf{N}$ to hold, gauge changes are a more general type of transformations than simple vector-field induced flows. Consequently, the Taylor-like expansion must be of a more general type. Indeed, for a given gauge change between $X$ and $Y$, the family of gauge changes $\phi_{X \rightarrow Y, \lambda}$ labeled by $\lambda$ is not always a group in $\lambda$, and this happens for instance if $[X,Y] \neq 0$ (See Ref.~\cite{Bruni-1997} for a graphic intuition). Although we could, for a fixed $\lambda = \lambda_{0}$, find $\xi$ such that Eq.~(\ref{knight}) takes a form like Eq.~(\ref{perturbation-tenseur}) up to a given order, for instance by fixing $\lambda_{0} = 1$, and choosing

\begin{equation}
\xi \equiv \xi_1 + \demi \xi_2 + \frac{1}{3!}\left(\xi_3+\frac{3}{2}[\xi_1,\xi_2]\right),
\end{equation}
this would mean that intermediate space-times are useless, and we would then ask Einstein equations to hold only for $\mathcal{P}_0(\mf{N})$ and $\mathcal{P}_{\lambda_0}(\mf{N})$. This would lead to equations in the perturbation variables that mix different orders. The resulting solution, for second order and above, would be very difficult to find. 

\subsection{Perturbed Einstein equations}

Instead, we prefer to use this more complicated but cleaner knight-diffeomorphism (Eq.~\ref{transforule}) to change gauge. It keeps the differential structure built on $\mf{N}$ and we additionally demand Einstein equations to be satisfied on each ${\cal P}_{\lambda}(\mathcal{N})$. This can be used to differentiate Einstein equations to first order w.r.t $\lambda$ and take the limit $\lambda \rightarrow 0$ in order to get a set of equations that formally take the form $\mathcal{E}_1[\delta^{(1)} g ,\delta^{(1)} T]=0$. Once solved for the solutions of the first-order Einstein equation, we can differentiate twice the Einstein equation w.r.t $\lambda$ and get an equation of the type  

\begin{equation} 
\mathcal{E}_2[\delta^{(2)} g ,\delta^{(2)} T]= \mathcal{S}[\delta^{(1)} g ,\delta^{(1)} T],
\end{equation}
where $S$ stands for a source term quadratic in the first-order variables (see
\cite{Bob} for a concrete example).

We see that the decomposition of perturbation variables in the form given by Eq.~(\ref{decomposition-ordre2}) will trigger a similarity between the equations, i.e. $\mathcal{E}_1$ and $ \mathcal{E}_2$ have the same form. Purely second-order perturbation variables satisfy the same equation as first-order perturbation variables do, but with a source term. With known sources and known solutions to the homogenous equation, the Green function method enables us to solve, at least formally, the second-order equations, and by recursion at any order. To summarize, the Taylor expansion ``taylorizes'' the process for solving the equations by dividing tasks among orders, since Einstein equations are satisfied order by order. 

\subsection{Gauge-invariant variables}

General covariance, i.e. the fact that physics should not depend on a
particular choice of coordinates is an incentive to work with gauge-invariant quantities. As we notice from Eq.~(\ref{transforule}), a tensor $T$ is gauge-invariant up to $n$-th order if it satisfies $\Lie_{\xi}\delta^{(r)}_X T=0$ for any vector field $\xi$ and any $r \leq n$, as can be deduced by recursion. A consequence of this strong condition is that a tensor is gauge-invariant up to order $n$ if and only if $T_0$ and all its perturbations of order lower than $n$ either vanish, or are constant scalars, or are combinations of Kronecker deltas with constant coefficients. Einstein equation is of the form $G-T=0$, and for this reason is totally gauge invariant. However, we cannot find non-trivial tensorial quantities (that is, different from $G-T$) gauge-invariant up to the order we intend to study perturbations, with which we could express the perturbed set of Einstein equations.

Consequently, we will lower our goal and we will build, by combinations of perturbed tensorial quantities, gauge-invariant variables. These combinations will not be the perturbation of an underlying tensor. This method will prove to be very conclusive since a general procedure exists for perturbations around FL. Eventually we shall identify observables among these gauge-invariant variables and the fact that they are not the perturbation of a tensor will not matter. It has to be emphasized that the transformation rules of these combinations are not intrinsic and cannot be deduced directly from the knight-diffeomorphism since they are not tensorial quantities. Instead, we have to form the combination before and after the gauge change in order to deduce their transformation rules.

We now summarize the standard way to build gauge-invariant variables. For
simplicity we consider only the scalar part of the gauge transformations,
since we will not consider vector modes in the metric and fluid perturbation
variables (again, this could be done, but would just obfuscate the
explanations). In the following, we split $\xi^{\mu}_r$ as
\begin{equation}
\xi^0_r=T^{(r)},\,\,\,\,\,\xi^{i}_r=\dhi L^{(r)},\,\,\,{\rm with}\,\,r =1,2.
\end{equation}

\subsection{First-order gauge-invariant variables}

In the subsequent work we present the transformation rules of perturbed
quantities in a simplified notation. Instead of writing
$W^{(r)}_Y=W^{(r)}_X+f\left(\xi_{1},..,\xi_{r}\right)$, in order to state that
the difference between the expression of the $r$-th order pertubed variable
$W$ in gauge $Y$ and in gauge $X$ is a function $f$ of the
knight-diffeomorphism fields $\xi_{1},...,\xi_{r}$, we prefer to write
$W^{(r)} \rightarrow W^{(r)} + f\left(\xi_{1},..,\xi_{r}\right)$. We remind
that the expressions of the fields $\left(\xi_{i}\right)_{1\leq i\leq r}$
necessary for the knight-diffeomorphism are expressed in function of the gauge
fields $X$ and $Y$ [see below Eq.~(\ref{knight})]. From the transformation rules~(\ref{transforule}) we deduce that the first-order perturbations of the metric tensor (\ref{metric}) transform as

\begin{eqnarray}
\Phi^{(1)} & \rightarrow & \Phi^{(1)} + T^{(1)'} + \HH T^{(1)}\\
B^{(1)} & \rightarrow & B^{(1)} -T^{(1)} + L^{(1)'}\\
\Psi^{(1)} & \rightarrow & \Psi^{(1)} -\HH T^{(1)}\\
E^{(1)} & \rightarrow & E^{(1)} + L^{(1)}\\
E^{(1)}_{ij} & \rightarrow & E^{(1)}_{ij},
\end{eqnarray}
while the scalar quantities related to matter transform as 
\begin{eqnarray}\label{Tfluide1}
\delta^{(1)} \rho & \rightarrow & \delta^{(1)} \rho + \bar{\rho}'T^{(1)} \nonumber\\
\delta^{(1)} P & \rightarrow & \delta^{(1)} P + \bar{P}'T^{(1)} \nonumber\\
v^{(1)} & \rightarrow & v^{(1)} - L^{(1)'}\\
\delta^{(1)} \pi^{ij} & \rightarrow& \delta^{(1)} \pi^{ij},
\end{eqnarray}
where a prime denotes a derivative w.r.t conformal time $\eta$, and where $\HH \equiv a'/a$.

From now on, we shall refer to these first-order transformation rules defined by $\xi_1$ as $\Tr_{\xi_1}(\Phi^{(1)}),\Tr_{\xi_1}(B^{(1)}),...$ or simply $\Tr(\Phi^{(1)}),\Tr(B^{(1)}),...$ For instance $\Tr(\Phi^{(1)})= \Phi^{(1)} +  T^{(1)'}+ \HH T^{(1)}$.

We first note that the first-order tensorial modes and the first-order
anisotropic stress are automatically gauge
invariant. For the other perturbation variables, which are not automatically gauge invariant, they are two ways to understand the procedure to build gauge-invariant combinations.
The first point of view in building gauge-invariant variables consists in finding a way to get rid of the undesired transformation rule. To do so, we remark that the combinations $B^{(1)}-E^{(1)'}$ and $-E^{(1)'}$ transform under a gauge change as $B^{(1)}-E^{(1)'} \rightarrow B^{(1)}-E^{(1)'} -T^{(1)} ,\,\,\,\, -E^{(1)} \rightarrow -E^{(1)} -L^{(1)}$.
We can use these combinations to add {\it ad hoc} compensating terms to $\Phi^{(1)}$ and $\Psi^{(1)}$ by defining
\begin{equation}
\hat{\Phi}^{(1)} \equiv \Phi^{(1)} + \left(B^{(1)}-E^{(1)'}\right)' + \HH \left(B^{(1)}-E^{(1)'}\right)
\end{equation}
\begin{equation}
\hat{\Psi}^{(1)}\equiv\Psi^{(1)} -\HH \left(B^{(1)}-E^{(1)'}\right).
\end{equation}
$\hat{\Phi}^{(1)}$ and $\hat{\Psi}^{(1)}$ are now gauge invariant, by construction.
This can also be understood, from a second point of view, as a gauge
transformation for $\Phi^{(1)}$ and $\Psi^{(1)}$ towards the Newtonian gauge
(NG) \cite{Mukhanov1992} defined by $\xi^{(1)}_{\rightarrow NG}$ decomposed in $T^{(1)}_{\rightarrow NG}= B^{(1)}-E^{(1)'} ,\,\,\, L^{(1)}_{\rightarrow NG}=-E^{(1)} $, which transforms the perturbation variables as
\begin{eqnarray}\label{transfo_order_1}
 B^{(1)} & \rightarrow & 0 \\
 E^{(1)} & \rightarrow & 0 \\
\Phi^{(1)} & \rightarrow & \hat{\Phi}^{(1)} \equiv \Phi^{(1)}_{NG}=\Phi^{(1)} + \HH\left(B^{(1)}-E^{(1)'}\right)+\left(B^{(1)}-E^{(1)'}\right)' \\
\Psi^{(1)} & \rightarrow & \hat{\Psi}^{(1)} \equiv \Psi^{(1)}_{NG}=\Psi^{(1)}- \HH\left(B^{(1)}-E^{(1)'}\right).
\end{eqnarray}
Similarly the gauge-invariant variables that would reduce to $\delta \rho$, $\delta P$  and $v$ are

\begin{eqnarray}\label{deffluideGI1}
\delta^{(1)} \hat{\rho}& \equiv &\delta^{(1)}_{NG} \rho= \delta^{(1)} \rho + \bar{\rho}' \left(B^{(1)}-E^{(1)}\right)'\nonumber\\
\delta^{(1)} \hat{P}& \equiv &\delta^{(1)}_{NG} P= \delta^{(1)} P + \bar{P}' \left(B^{(1)}-E^{(1)'}\right)\nonumber\\
\hat{v}^{(1)}& \equiv &v^{(1)}_{NG}= v^{(1)} + E^{(1)'}\nonumber\\
\hat{\pi}^{ij(1)} & \equiv & \delta^{(1)}_{NG} \pi^{ij} = \delta^{(1)} \pi^{ij}.
\end{eqnarray}

Since we have ignored the vector gauge degrees of freedom, $B^{(1)}$ and $E^{(1)}$ are the two gauge variant variables of the metric perturbation while $\hat{\Phi}^{(1)}$ and $\hat{\Psi}^{(1)}$ are the gauge-invariant part. As mentionned before, we then force the gauge-invariant variables in the perturbed metric by replacing $\Phi^{(1)}$ with $\hat{\Phi}^{(1)} - \HH\left(B^{(1)}-E^{(1)'}\right)+\left(B^{(1)}-E^{(1)'}\right)'$ and applying similar procedures for $\Psi^{(1)}$, $\delta^{(1)} \rho$ and $\delta^{(1)} P$. When developping Einstein equations, we know that general covariance will eventually keep only gauge-invariant terms. Thus, we can either do a full calculation and witness the terms involving the degrees of freedom $B^{(1)}$ and $E^{(1)}$ disappear, or perform the calculations with $B^{(1)}$ and $E^{(1)}$ set to zero and obtain the perturbed Einstein equations only in function of gauge-invariant variables. The latter simplifies the computation, which is useful when going to higher orders. 
The advantage of the second point of view, is that the addition of the compensating terms of the first point of view can be seen as a first-order gauge change towards the Newtonian gauge with $\xi^{(1)}_{\rightarrow NG}$ (decomposed as $T^{(1)}_{\rightarrow NG}$ and $L^{(1)}_{\rightarrow NG}$). These enables us to decompose the perturbed metric in a gauge-invariant part and a gauge variant part as

\begin{equation}
\delta^{(1)}g \equiv \delta^{(1)}\tilde{g} + \Lie_{-\xi^{(1)}_{\rightarrow NG}} \bar{g},
\end{equation}
as it can be seen from the transformation rules under a gauge change characterised by $\xi_1$
\begin{eqnarray}
\delta^{(1)}\tilde{g} &\rightarrow&  \delta^{(1)}\tilde{g},\nonumber\\
-\xi^{(1)}_{\rightarrow NG} &\rightarrow& -\xi^{(1)}_{\rightarrow NG} + \xi_1.
\end{eqnarray}

This property which is not general but happens to hold in the case of cosmological perturbation (i.e. around  FL metric) is the key to extend this construction to second order.

It should be noted that this procedure, although achieved by defining gauge
invariant variables which reduce to the perturbation variables in the
Newtonian gauge, can be extended to other types of gauge-invariant variables
which reduce to perturbation variables in another gauge. For instance, we can
use the transformation properties of $\Psi^{(1)}$ and $E^{(1)}$ to add the compensating
terms to $\Phi^{(1)}$, $B^{(1)}$ and other variables. The transformation rules
$\Psi^{(1)} / \HH \rightarrow \Psi^{(1)} / \HH  -T^{(1)} ,\,\,\,-E^{(1)} \rightarrow -E^{(1)} -L^{(1)}$
make it straightforward to build these compensating terms. We need to define
$\xi^{(1)}_{\rightarrow FG}$ decomposed with $T^{(1)}_{\rightarrow FG}= \Psi^{(1)}
/ \HH ,\,\,\, L^{(1)}_{\rightarrow FG}=-E^{(1)} $. The gauge-invariant variables
defined with this procedure reduce to the perturbation variables
in the flat gauge ($E^{(1)}=0$, $\Psi^{(1)}=0$) and are

\begin{equation}
\tilde{B}\equiv B_{FG}= B^{(1)}-\frac{\Psi^{(1)}}{\HH}-E^{(1)'} ,\,\,\, \tilde{\Phi}^{(1)} \equiv \Phi^{(1)}_{FG} =\Phi^{(1)} + \Psi^{(1)} + \left(\frac{\Psi^{(1)}}{\HH}\right)'.
\end{equation}

\subsection{Second-order gauge-invariant variables}\label{defGIV}

For second-order perturbations, Eq.~(\ref{transforule}) gives the following transformation rules

\begin{eqnarray}\label{transfo_order_2}
\Phi^{(2)} & \rightarrow & \Phi^{(2)} + T'^{(2)} + \HH T^{(2)} + S_{\Phi}\nonumber\\
B^{(2)} & \rightarrow & B^{(2)} -T^{(2)} + L'^{(2)} + S_{B}\nonumber\\
\Psi^{(2)} & \rightarrow & \Psi^{(2)} -\HH T^{(2)} + S_{\Psi}\nonumber\\
E^{(2)} & \rightarrow & E^{(2)} + L^{(2)} + S_{E}\nonumber\\
E^{(2)}_{ij} & \rightarrow & E^{(2)}_{ij}+ {S_{E}}_{ij}\nonumber\\
\delta^{(2)} \rho & \rightarrow & \delta^{(2)} \rho + \bar{\rho}'T^{(2)} + S_{\rho}\nonumber\\
\delta^{(2)} P & \rightarrow & \delta^{(2)} P + \bar{P}'T^{(2)} + S_{P}\nonumber\\
v^{(2)} & \rightarrow & v^{(2)} -L^{(2)'} + S_{v}\nonumber\\
\pi^{ij(2)} & \rightarrow& \pi^{ij(2)} + 2 T^{(1)}\left(\pi^{ij(1)}\right)' +2 \dhk L^{(1)} \dbk \pi^{ij(1)} \nonumber\\
&& \quad - 2 \pi^{ik(1)} \dbk \dhj L^{(1)} - 2 \pi^{jk(1)} \dbk \dhi L^{(1)},
\end{eqnarray}
where the source terms are quadratic in the first-order variables
$T^{(1)},L^{(1)},\Phi^{(1)},\Psi^{(1)}$. We collect the expressions of these terms in~\ref{app_sources}.
In the rest of this paper, we shall refer to these second-order transformation rules associated with $(\xi)\equiv (\xi_1,\xi_2)$ as $\Tr_{(\xi)}(\Phi^{(2)}),\Tr_{(\xi)}(B^{(2)}),...$ or simply $\Tr(\Phi^{(2)}),\Tr(B^{(2)}),...$.
These transformation rules are much more complicated than their first-order
counterparts. However, the combination defined by $F \equiv \delta^{(2)}g + 2
\Lie_{\xi^{(1)}_{\rightarrow NG}} \delta^{(1)}g + \Lie_{\xi^{(1)}_{\rightarrow
    NG}}^2 \bar{g} $ enjoys the simple transformation rule $F \rightarrow  F +
\Lie_{\xi_2 + [\xi^{(1)}_{\rightarrow NG}, \xi_1 ]} \bar{g}$ under a gauge change
defined by $\xi_2$ and $\xi_1$ (see Ref.~\cite{Nakamura-2006}). As a result,
its transformation rule mimics the one of first-order pertubations under a
gauge change. This means that if we decompose $F$ in the same way as we did
for the metric the metric with 
\begin{eqnarray}
\Phi_F &\equiv&\Phi^{(2)}+S_{\Phi}(\xi^{(1)}_{\rightarrow NG})\nonumber\\ 
\Psi_F &\equiv& \Psi^{(2)}+S_{\Psi}(\xi^{(1)}_{\rightarrow NG}) \nonumber\\ 
B_F &\equiv& B^{(2)}+S_{B}(\xi^{(1)}_{\rightarrow NG}) \nonumber\\ 
E_F &\equiv& E^{(2)}+S_{E}(\xi^{(1)}_{\rightarrow NG})\nonumber\\  
E_{Fij} &\equiv& E^{(2)}_{ij}+{S_{E}}_{ij}(\xi^{(1)}_{\rightarrow NG}),  
\end{eqnarray}
then the transformation rules for these quantities will be similar to those of
Eq.~(\ref{transfo_order_1}), but with the vector $\xi_2 + [\xi_{\rightarrow
  NG}, \xi_1 ]$ instead of $\xi_1$. Consequently, we shall use the same combinations in order to construct gauge-invariant variables which are
\begin{eqnarray}
\hat{\Phi}^{(2)} &\equiv& \Phi_F + \left(B_F-E_F'\right)' + \HH\left(B_F-E_F'\right)\nonumber\\
\hat{\Psi}^{(2)} &\equiv& \Psi_F -\HH\left(B_F-E_F'\right)\nonumber\\
\hat{E}_{ij}^{(2)} & \equiv & E_{Fij}.
\end{eqnarray}

As for the first order, this addition of compensating terms can be understood,
from the second point of view, as a defining the gauge-invariant variables as
the perturbation variables in a given gauge. In our case it is the Newtonian
gauge since it transforms $B$ and $E$ into a null value. This transformation is defined by
$\xi^{(2)}_{\rightarrow NG}$ that we decompose in 
\begin{eqnarray}
T^{(2)}_{\rightarrow  NG}&=&B^{(2)}-E^{'(2)} +
S_{B}^{(2)}\left(\xi^{(1)}_{\rightarrow
    NG}\right)-S_{E}^{'(2)}\left(\xi^{(1)}_{\rightarrow  NG}\right)\nonumber\\
L^{(2)}_{\rightarrow NG}&=&-E^{(2)} -S_{E}^{(2)}\left(\xi^{(1)}_{\rightarrow  NG}\right).
\end{eqnarray}
The second-order gauge-invariant variables can be seen as

\begin{eqnarray}
\hat{\Phi}^{(2)} &\equiv& \delta^{(2)}_{NG} \Phi\nonumber\\
\hat{\Psi}^{(2)} &\equiv& \delta^{(2)}_{NG}\Psi\nonumber\\
\hat{E}^{(2)}_{ij} & \equiv & \delta^{(2)}_{NG} E_{ij}
\end{eqnarray} 
\begin{eqnarray}\label{deffluideGI2}
\delta^{(2)} \hat{\rho}&\equiv&\delta^{(2)}_{NG} \rho\nonumber\\
\delta^{(2)} \hat{P}&\equiv&\delta^{(2)}_{NG} P\nonumber\\
\hat{v}^{(2)}& \equiv& v^{(2)}_{NG}\nonumber\\
\hat{\pi}^{ij(2)} & \equiv & \delta^{(2)}_{NG} \pi^{ij}.
\end{eqnarray}
where the index $NG$ indicates that we transformed the quantity with the
formula (\ref{transforule}), with the vector fields $\xi^{(1)}_{\rightarrow
  NG}$ and $\xi^{(2)}_{\rightarrow NG}$ defined above. This means that we have split the second-order metric according to

\begin{equation}
\delta^{(2)}g = \delta^{(2)}\tilde{g} + \Lie_{-\xi^{(2)}_{\rightarrow NG}} \bar{g}+2\Lie_{-\xi^{(1)}_{\rightarrow NG}} \delta^{(1)} g-\Lie^{2}_{-\xi^{(1)}_{\rightarrow NG}} \bar{g}
\end{equation}
where $\delta^{(2)}\tilde{g}$  is the gauge-invariant part and
$-\xi^{(2)}_{\rightarrow NG}$ the gauge variant part, as it can be seen from the transformation rules under a gauge change characterised by $(\xi_1,\xi_2)$
\begin{eqnarray}
\delta^{(2)}\tilde{g} &\rightarrow&  \delta^{(2)}\tilde{g},\nonumber\\
-\xi^{(2)}_{\rightarrow NG} &\rightarrow& -\xi^{(2)}_{\rightarrow NG} + \xi_2 + [\xi^{(1)}_{\rightarrow NG}, \xi_1 ].
\end{eqnarray}

As for the first order, we can choose other types of combinations, for
instance those which are equivalent to setting the gauge as being flat, by
using this procedure. In this case, the vector field $\xi^{(2)}_{\rightarrow
  FG}$ is decomposed in 

\begin{equation}
T^{(2)}_{\rightarrow FG}=\frac{\Psi^{(2)}}{\HH} + \frac{1}{\HH}S_{\Psi}^{(2)}\left(\xi^{(1)}_{\rightarrow  FG}\right),\,\,\,L^{(2)}_{\rightarrow
  FG}=-E^{(2)} -S_{E}^{(2)}\left(\xi^{(1)}_{\rightarrow  FG}\right).
\end{equation}

It should also be mentioned that the existence of an inverse Laplacian
$\Delta^{-1}$ of the background space-time, i.e. a corresponding Green function with boundary conditions, is required for all this procedure. In other words, when working in Fourier space, all our conclusions will be valid only for modes which do not belong to the Kernel of $\Delta$. 

\section{Gauge transformation of the distribution function}

\subsection{pre-Riemannian distribution function}

So far, we have set up the mathematical framework to identify points between
the background space-time and the perturbed space-times through a gauge
field $X$. This enabled us to define the perturbation of tensors and to
calculate their transformation properties under a gauge transformation.
However this only allows to perform a fluid treatment of the radiation. In
the statistical description for a set of particles, we assume that each
particle has a given impulsion $p^{\mu}$ and is located at a given position
\cite{Hillery}. The equations then have to describe the phase space distribution of the particles. If the number of particles is high enough, we can define a
probability density, the distribution function, of finding a particle in an
infinitesimal volume of the phase space. Now, let us focus our attention on
this distribution function. The distribution function is a function of the
point considered (i.e. its coordinates $x^{\mu}$), and also a function of the
tangent space at this point whose coordinate we label by
$p^{\nu}\partial_{\nu}$. There is no special reason for this function to be
linear in $p^{\nu}\partial_{\nu}$, but we can expand it, without any loss of generality, in power series of tensors according to 

\begin{equation}
f\left(x^{\nu},p^{\nu} \right) = \sum_k {\mathcal F}_{\mu_1..\mu_k}(x^{\nu})p^{\mu_1}...p^{\mu_k}.
\end{equation}
The distribution function is then decomposed as the sum of all the multipoles ${\mathcal
   F}_{\mu_1..\mu_k}$ evaluated in a particular point of the tangent space.
 From the previous section we know the transformation rules for these tensorial
 quantities, thus $f$ transforms according to

\begin{equation}
\Tr_{(\xi)}\left[ f\left(x^{\nu},p^{\nu}  \right)\right]  \equiv \sum_k \Tr_{(\xi)}\left[{\mathcal F}_{\mu_1..\mu_k}(x^{\nu})\right]p^{\mu_1}...p^{\mu_k},
\end{equation}
where $\Tr_{(\xi)}$ refers to the knight-diffeomorphism with the set of vectors
$(\xi_1,\xi_2,...)$.

As we do not necessarily want to refer explicitly to the decomposition in multipoles, we
use the fact that for any vector $\xi = \xi^{\mu}\partial_{\mu}$, which defines a flow on the background space-time $\mathcal{P}_{0}(\mf{N})$, we can define an induced flow (a natural lift) on the vector tangent bundle $T\mathcal{P}_{0}(\mf{N})$ directed by the vector field $T\xi =
 \left[\xi^{\mu}\partial_{\mu},p^{\nu}(\partial_{\nu}\xi^{\mu})\frac{\partial}{\partial p^{\mu}}\right] $. This implies the useful property
 
\begin{equation}
\Lie_{\xi}\left({\mathcal F}_{\mu_1..\mu_p}\right)p^{\mu_1}..p^{\mu_p} = \Lie_{T\xi}\left({\mathcal F}_{\mu_1..\mu_p}p^{\mu_1}..p^{\mu_p}\right).
\end{equation}
With this definition, we can rewrite the transformation rule for $f$ as

\begin{equation}\label{Txi}
\Tr_{(\xi)}\left[ f\left(x^{\nu},p^{\nu} \right)\right]  = \Tr_{(T\xi)}\left[ f\left(x^{\nu},p^{\nu} \right)\right],
\end{equation} 
where now $\Tr_{(T \xi)}$ refers to the knight-diffeomorphism with the set of
vectors $(T\xi_1,T\xi_2,...)$.

The evolution of the distribution function is dictated by the
Boltzmann equation $\frac{\dd f}{ \dd \eta}=C[f]$, where the r.h.s is the
collision term which encodes the local physics. This collision term can be easily expressed in the local Minkowskian frame defined by a tetrad fields $e_a$, from known particles physics. For this reason, the framework developed to define gauge
transformations for a general manifold has to be extended to the case of
Riemannian manifold. Instead of using the coordinates basis $\partial_{\mu}$ to
express a vector of tangent space as $V=p^{\mu}\partial_{\mu}$, we prefer to use the tetrads basis $e_a$ and write
$V=\pi^a e_a$. In terms of coordinates, this means that the distribution function is a function
of $x^{\mu}$ and $\pi^a$. When expressing the physics with the tetrad fields, the metric is not just one of the many tensors of the theory whose properties under a gauge transformation we need to know, but
rather a central feature of the manifold, since it determinates the tetrads (up to a Lorentz tranformation) required to express the distribution function. As the metric is a tensor, and as the tetrads are defined according to the metric, the extension is inherited from the previous section.    

\subsection{Tetrads}

\subsubsection{Definitions}

On each slice $\mathcal{P}_{\lambda}(\mf{N})$, we should have four vector
fields\footnote{The fifth direction which arises from the extension of the
  manifold from ${\cal M}$ to ${\cal N}$ is ignored as the component of any
  tensor is required to vanish in this direction. We thus consider the tangent
  space at each point of ${\cal N}$ as being four-dimensional.} (and their associated 1-form fields) labeled by $a=0,1,2,3$,
which satisfy the normalization conditions

\begin{equation}\label{deftetrad}
e^{\mu}_a e^{\nu}_b g_{\mu\nu}= \eta_{ab},\,\,\,\,\,\,e_{\mu}^a e_{\nu}^b g^{\mu\nu}= \eta^{ab}.
\end{equation} 
With these notations, indices $a,b,c..$ are raised and lowered with $\eta_{ab}$.

With the formalism developed for tensors, we carry this tetrad field
onto the background space-time using a gauge field $X$ with

\begin{eqnarray}
&&e_{a,X}^{\mu} \equiv \phi^{\star}_{\lambda,X}(e_a)= \sum_{k=0}^{k=\infty}\frac{\lambda^k}{k!}\Lie_X^k e_a\nonumber\\
&&\delta^{(n)}_X e_a \equiv \Lie_X^n e_a \Big|_{\mathcal{P}_0(\mf{N})},\,\,\,\,\,\bar{e}_a \equiv \delta^{(0)}_X e_a,
\end{eqnarray}  
and similar formulas for $e^a$.

As $\eb_a$ is a basis of the tangent space on the background space-time (and
$\eb^a$ a basis of its dual space),
$e^{\mu}_{a,X}$ and $e_{\mu,X}^{a}$ can be expressed in the
generic form
\begin{equation}\label{defRS}
e_{a,X} = R^{\,\,b}_{a,X}\eb_b,\,\,\,e^b_{X} = \eb^a S^{\,\,b}_{a,X},\,\,R^{\,\,c}_{a,X} S^{\,\,b}_{c,X}=S^{\,\,c}_{a,X}
R^{\,\,b}_{c,X}=\delta^{b}_{\,a},
\end{equation}
where, 
\begin{eqnarray}\label{Devsimplifies}
R_{ab,X}  &\equiv& \sum_k\frac{\lambda^k}{k!}R^{(k)}_{ab,X}\nonumber\\
S_{ab,X} &\equiv& \sum_k \frac{\lambda^k}{k!}S^{(k)}_{ab,X}.
\end{eqnarray}
Order by order, this reads
\begin{equation}
\delta^{(n)}_X e_a =R^{(n)b}_{a,X} \eb_b,\,\,\,\delta^{(n)}_X e^b = \eb^a
S^{(n)b}_{a,X}. 
\end{equation}

\subsubsection{Normalization condition}\label{sec:choicetetrad}

Tetrads are four vector fields which satisfy Eq.~(\ref{deftetrad})
and are thus related to the metric. Consequently, the perturbations of the
tetrad defined above are partly related to the perturbations of the metric.
When pulled back to the background space-time, Eq.~(\ref{deftetrad}) implies

\begin{eqnarray}
\phi^{\star}_{\lambda,X}(\eta_{ab})=\eta_{ab} &=&\phi^{\star}_{\lambda,X}(e^{\mu}_a e^{\nu}_b g_{\mu\nu})\nonumber\\
&=&\phi^{\star}_{\lambda,X}(e^{\mu}_a) \phi^{\star}_{\lambda,X}(e^{\nu}_b) \phi^{\star}_{\lambda,X}(g_{\mu\nu}).
\end{eqnarray}
Identifying order by order (in terms of $\lambda$) we get in particular for
the first and second orders
\begin{eqnarray}\label{Constraints}
\eb_b.\delta^{(1)}_X e_a + \eb_a.\delta^{(1)}_X e_b + \delta^{(1)}_X
g(\eb_a,\eb_b)&=&0\nonumber\\
 \eb_b.\delta^{(2)}_X e_a + \eb_a.\delta^{(2)}_X e_b + \delta^{(2)}_X
g(\eb_a,\eb_b)&&\\
+ \delta^{(1)}_X e_b . \delta^{(1)}_X e_a  +\delta^{(1)}_X
g\left(\delta^{(1)}_X e_a,\eb_b\right)+ \delta^{(1)}_X
g\left(\eb_a,\delta^{(1)}_X e_b\right)&=&0,\nonumber
\end{eqnarray}
where a dot product stands for $\bar{g}\left(\_\,,\_\right)$.
From the constraints (\ref{Constraints}), we can determine the symmetric part
of $R^{(n)}_{ab}$ as
\begin{eqnarray}\label{Rfdeg}
R^{(1)}_{(ab),X}&=&-\demi \delta^{(1)}_X g(\eb_a,\eb_b)\\
R^{(2)}_{(ab),X}&=&-\demi \delta^{(2)}_X g(\eb_a,\eb_b) -\delta^{(1)}_X
g\left(R^{(1)}_{ac,X}\eb^c,\eb_b\right) \nonumber\\
&& \quad- \delta^{(1)}_X g\left(\eb_a,R^{(1)}_{bc,X} \eb^c\right)-
R^{(1)c}_{a,X}R^{(1)}_{bc,X}, 
\end{eqnarray}
which are related to the components of the inverse by
\begin{eqnarray}
S^{(1)}_{ab,X} &=& -R^{(1)}_{ab,X}\\
S^{(2)}_{ab,X} &=& -R^{(2)}_{ab,X}+ 2R^{(1)c}_{a,X}R^{(1)}_{cb,X}.
\end{eqnarray}

The antisymmetric part, $R_{[ab],X}$, still remains to be chosen as it corresponds
to the Lorentz transformation freedom (boost and rotation), which is allowed by the
definition~(\ref{deftetrad}).
A first and easy choice would be $R^{(n)}_{[ab],X}=0$ for any $n$. However, as
mentioned above, we eventually want to decompose a vector $p^{\mu}\partial_{\mu}$ on tangent space as

\begin{equation}\label{ptopi}
p^{\mu}\partial_{\mu}= \pi^a e_a =\pi^a e_a^{\mu}\partial_{\mu},
\end{equation}
and identify $\pi^0$ with the energy and $\pi^i$ with the momentum (although
conserved quantities are generally ill-defined in general relativity, energy
and momentum can be defined when performing perturbations around a maximally symmetric
background \cite{Nathalie} as it is the case here). When working with coordinates, we
want to express physical quantities, as measured by comoving observers, i.e. observers of constant spatial coordinates, whose motion is defined by the 1-form $(d\eta)_{\mu}$
\cite{Gourgoulhon}.  We thus require $(e^0)_{\mu} \sim (d\eta)_{\mu} $, which is equivalent to choose $S^{(n)}_{a_i 0,X}=0$
for any $n$, where $a_i=1,2,3$. This choice allows us to fix the
boost in $S^{(n)}$ by imposing the condition $S^{(n)}_{[a_i 0],X}=
-S^{(n)}_{[0 a_i],X}=-S^{(n)}_{(a_i 0),X}$. As Eq.~(\ref{defRS}) implies that for any $n$
\begin{equation}\label{recursion}
S_{ac}^{(n)}+R_{ac}^{(n)}+\prod_{\begin{array}{c}
             \{\,p+q=n,\\ \qquad p\ge1,\,q\ge1 \}
             \end{array}} S_{a}^{(p)b}R_{bc}^{(q)}\frac{n!}{p!q!}=0,
\end{equation}
it can be checked by recursion that this implies 
\begin{equation}
R^{(n)}_{[a_i 0],X}= -R^{(n)}_{[0  a_i],X}=-R^{(n)}_{(a_i 0),X}.
\end{equation}
We also fix the rotation by requiring $S^{(n)}_{[a_i a_j],X}=0$, and it can be checked similarly, by recursion on Eq.~(\ref{recursion}), that this implies $R^{(n)}_{[a_i a_j],X}=0$. 

\subsection{Gauge transformation of tetrads}

Under a gauge transformation, we can deduce the transformation
properties of the tetrad from those of the perturbed metric. For simplicity, we restricted to
scalar and tensor perturbations, but this is completely general and can be easily
extended to include vectors. 
In the FL case, we use a natural background tetrad associated to Cartesian coordinates $\eb_0 =
\left(\partial_{\eta}\right)/a,\,\,\eb_{b_i} = \left(\partial_{i}\right)/a $,
in order to evaluate Eq.~(\ref{Rfdeg}). 
The notation $b_i$ refers to Lorentz (SO(1,3)) indices running from $1$ to $3$, whereas $i$ is a coordinate index running from $1$ to $3$.
When uselessly obfuscating the explanation, we will not make the distinction
and change $b_i$ for $i$.
We report the detailed expressions for the transformation of the tetrads for
the first and second orders in \ref{app_TRS}.

\section{Distribution function}

Now that the transformation properties of the tetrads are known, we turn to the general transformation of a distribution function $f(x^{\mu},\pi^a)$.

\subsection{Multipolar expansion}\label{multipolar}

Any function $f(x^{\mu},\pi^{a})$ can be expanded in symmetric trace free
multipoles as \cite{Thorne} 
\begin{equation}
f(x^{\mu},\pi^{a})= \sum_p F_p(x^{\mu},\pi^{a}) 
\end{equation}
with 
\begin{eqnarray}\label{decmultipoles}
F_p(x^{\nu},\pi^{a}) & \equiv & {\mathcal F}_{\mu_1..\mu_p}(x^{\nu})p^{\mu_1}..p^{\mu_p}\nonumber\\
& = &\left[{\mathcal F}_{\mu_1..\mu_p}(x^{\nu})e_{a_1}^{\mu_1}..e_{a_p}^{\mu_p}\right]\pi^{a_1}..\pi^{a_p}\nonumber\\
&\equiv& F_{a_1..a_p}(x^{\nu})\pi^{a_1}..\pi^{a_p}.
\end{eqnarray}

We do not need any additional identification procedure for the tangent spaces through a gauge field, in order to identify
  points of the tangent space of the slices $T{\mathcal P}_{\lambda}({\mathcal
    N})$. Indeed, once the metric and a gauge field $X$ are chosen, there exists a natural
  identification with the tetrad fields. First, and as mentioned before, we identify the points of ${\cal
    N}$ which lie on the same integral curves of $X$, that is, we identify a point $P \in{\cal P}_{0}({\cal N})$ and $\Phi_{\lambda,X}(P) \in {\cal
    P}_{\lambda}({\cal N})$. Then, we identify vectors of their respective tangent
  spaces, if the coordinates of these vectors in their respective local tetrad
  frames $\bar{e}_a$ and $e_a$, are the same. To be short, we identify $\pi^a
  e_a$ and $\pi^a \bar{e}_a$. As a consequence, for any given set
  $\{a_1,...,a_p\}$, the function $F_{a_1..a_p}(x^{\nu})$ is a scalar field.  $F_{a_1..a_p}(x^{\nu})$ is then pulled back on the
background space-time using the gauge field $X$, and we define in this way perturbations

\begin{equation}
\Phi^{\star}_{\lambda,X}\left[F_{a_1..a_p}(x^{\nu})\right]\equiv F_{X,a_1..a_p}(x^{\nu})\equiv\sum_{\lambda}\frac{\lambda^n}{n!} \delta^{(n)}_X F_{a_1..a_p}(x^{\nu}),
\end{equation}
and
\begin{equation}
F_{p,X}(x^{\nu},\pi^{a}) \equiv F_{X,a_1...a_p}(x^{\nu})\pi^{a_1}...\pi^{a_p}.
\end{equation} 
This perturbation scheme induces a perturbation procedure for the distribution
function $f$ as

\begin{eqnarray}\label{fXdev}
f_X(x^{\nu},\pi^a) &\equiv & \sum_{n}
\frac{\lambda^n}{n!} \delta^{(n)}_X f(x^{\nu},\pi^a),\nonumber\\
\delta^{(n)}_X f(x^{\mu},\pi^{a}) &\equiv& \sum_p  \delta^{(n)}_X F_{a_1...a_p}(x^{\nu})\pi^{a_1}...\pi^{a_p}.
\end{eqnarray}
It is essential to stress that $\pi^a$ is not a perturbed quantity, it is a
coordinate of the locally Minkowskian tangent space. However, the tetrad
field allows us to see $p^{\mu}$ as a perturbed vector since $p^{\mu}(\pi^a)
= e_a^{\mu} \pi^a$. In other words, for a given $\pi^a$, there is an associated vector
whose order by order perturbation in a given gauge $X$ is given by
$p^{\mu(n)}_{X} \equiv e_{a,X}^{\mu(n)} \pi^a$. 

\subsection{Gauge transformation: general case}\label{sec:generalcase}

We can deduce the transformation rule under a gauge change directly on the form~(\ref{decmultipoles}), pulled back to the background space-time,  
\begin{equation}\label{Tbrute}
\Tr\left[f_{X}(x^{\nu},\pi^{a})\right]\equiv \sum_p \Tr\left[{\mathcal F}_{X,\mu_1...\mu_p}(x^{\nu})\right]\Tr\left(e_{a_1,X}^{\mu_1}\right)...\Tr\left(e_{a_p,X}^{\mu_p}\right)\pi^{a_1}...\pi^{a_p}.
\end{equation}
The first factor in this expression is tensorial. Exactly as for the pre-Riemannian case, its transformation rule
is dictated by the knight-diffeomorphism, whereas we get the transformation
rules of the tetrads from Eqs.~(\ref{Ttetrads1}) and Eqs.~(\ref{Ttetrads2}). As we
do not necessarily want to refer explicitly to the multipole expansion, the first factor
is rewritten by considering $f$ as a function of $p^{\mu}$ using $\pi^a =
e^a_{\nu,X} p^{\nu}$, and applying Eq.~(\ref{Txi}). We then have to consider
the resulting distribution function as a function of $\pi^a$, knowing that the inversion is now given by
 $p^{\mu}(\pi^a)=\Tr(e^{\mu}_a) \pi^a$. This will account for $\Tr\left(e_{a_1,X}^{\mu_1}\right)$ in Eq.~(\ref{Tbrute}). In a compact form it reads

\begin{equation}\label{Tcompacte}
\Tr\left[f_X(x^{\nu},\pi^a)\right]=
     \Tr_{(T\xi)} \left\{f_X\left[x^{\nu},e^a_{\mu}p^{\mu}\right]\right\}\Big|_{p^{\mu}=\Tr(e_b^{\mu})\pi^b}.
\end{equation}

To obtain an order by order formula, we explicit these three steps using a Taylor expansion. First, we use that
\begin{equation}
f_X (x^{\nu},\pi^a)=\left[\exp\left(\eb^b_{\mu}p^{\mu}S_{b,X}^{\,\,\,a}\dsd{}{\pi^a}\right)f_X\right] (x^{\nu},\eb^b_{\mu}p^{\mu})\equiv g_X(x^{\nu},p^{\mu}),
\end{equation}
in order to consider $f$ as a function of $p^{\mu}$. We then Taylor expand back the result of the knight-diffeomorphism in order to read the result as a function of $\pi^a$,

\begin{equation}
\Tr\left[f_X (x^{\nu},\pi^a)\right]=\left[\exp\left(\eb_b^{\mu}\pi^a \Tr \left(R_{a,X}^{\,\,\,b}\right)\dsd{}{p^{\mu}}\right)\Tr_{(T\xi)} \left(g_X\right)\right](x^{\nu},\eb^{\mu}_a\pi^a).
\end{equation}

The derivatives in the previous expressions have to be ordered on the right
in each term of the expansion in power series of the exponential. When identifying order by
order, we need to take into account the expansion in $R_{ab}$ and $S_{ab}$, in
the exponentials and also in the knight-diffeomorphism. 

We have provided the general transformation rules for the distribution
function and we will specify now the transformation properties of the first-
and second-order distribution function.

\subsection{The mass shell}\label{mass_shell}

The transformation properties of $\delta^{(n)}_X
  e_{a}^{\mu}$  have been chosen so that, in the special case of
$f\equiv g_{\mu \nu}p^{\mu}p^{\nu}=g_{\mu \nu}e^{\mu}_a e^{\nu}_b \pi^a
\pi^b =\pi_a\pi^a$, it remains unchanged under a gauge transformation,
i.e. $\Tr(\pi^a \pi_a)=\pi^a \pi_a$. Since the tetrads must satisfy Eq.~(\ref{deftetrad}), then
$\delta_X^{(n)} f=0$ for $n \ge 1$, and it implies this property trivially. As a consequence, any
function of the form $\delta(\pi_a\pi^a-m^2)f(x^{\mu},\pi^a)$ transforms as
$\delta(\pi_a\pi^a-m^2)\Tr\left[f(x^{\mu},\pi^a)\right]$, where $m^2$ is the
mass of the particles described by the distribution function. In other words, the
transformation of the distribution function remains on the mass shell, as it has
been already mentionned in Ref.~\cite{Ruth}. We will make use of this
property when computing the transformation rules of the distribution function.

\section{Application to the perturbation of the Boltzmann equation for radiation}

The formalism developed in the previous section is general. We will now apply it to the particular FL case, and from now on we will also focus on the
radiation case, that is the case where $m^2=0$. For the first and the second order, we will present the transformation rules of the
distribution function for radiation, and build a gauge-invariant distribution function as
well as a gauge-invariant brightness. We will then write the evolution equation of this
gauge-invariant brightness in the case where the photon travels
freely through space-time without being affected by diffusion processes.
This is obtained using the collisionless Boltzmann equation

\begin{equation}
\frac{d f}{d \eta}=\dsd{f}{\eta}+\dsd{f}{x^i}\dsd{x^i}{\eta}+\dsd{f}{\pi^0}\dsd{\pi^0}{\eta}+\dsd{f}{n^i}\dsd{n^i}{\eta}=0,
\end{equation}
where $n^i \equiv \pi^i / \pi^0 $, from which we will extract the background, the first- and the second-order
equations after having pulled it back to the background space-time. In order to do so, we need to know $\dsd{\pi^0}{\eta}$ and
$\dsd{\pi^i}{\eta}$. By considering $p^{\mu}$ as a perturbed vector, as
mentionned in \S~(\ref{multipolar}), theses can be expressed from the geodesic equation
\begin{equation}
p^0 \frac{d p^{\mu}}{d \eta} = - \Gamma^{\mu}_{\nu \sigma}p^{\nu}p^{\sigma}
\end{equation}
that we pull back to the background space-time in order to extract order by order equations. Similarly,
$\dsd{x^i}{\eta}$ is given by the order by order expressions of $p^0 \dsd{x^i}{\eta}= p^i$, when pulled back to the background space-time.

At the background level, space is homogeneous and isotropic. Consequently, the distribution function depends neither on the direction $n^i$ of the photon nor on the position in space $x^i$. It only depends on $\pi^0$ and $\eta$, which
implies that $\dsd{\bar{f}}{n^i}=\dsd{\bar{f}}{x^i}=0$. Since the background geodesic deviation equation implies $\dsd{\pi^0}{\eta}=-\HH\pi^0$, the
collisionless Boltzmann equation reads at the background level

\begin{equation}\label{Boltzmann0}
\dsd{\bar{f}}{\eta}\Big|_{\pi}-\HH \pi^0 \dsd{\bar{f}}{\pi^0}=0.
\end{equation}

\subsection{Gauge transformation at first order}

In order to better understand the seemingly heavy but powerful formalism of \S~\ref{sec:generalcase}, let us apply it
to the first-order gauge transformation of the photon distribution function
$f$ in the Boltzmann equation. In this case, Eq.~(\ref{Tcompacte}) for
$\xi_1=(T,L)$ leads to

\begin{eqnarray}
&&\Tr\left[ \delta(\pi_c\pi^c)\delta^{(1)}_X f\right]=\nonumber\\
&& \qquad \delta(\pi_c\pi^c)\Big\{
\Lie_{T \xi_1}\left[\bar{f}(x^{\nu},a p^{\mu}) \right] + \left[\Tr\left(R^{(1)b}_{a,X}\right)+S^{(1)b}_{a,X}\right]\pi^a \dsd{\bar{f}}{\pi^b}\Big\}.
\end{eqnarray}

The expressions of $R_{a,X}^{\,\,\,b}$ and $S_{a,X}^{\,\,\,b}$, and their
transformation rules for the FL case, are given in \ref{app_TRS}. 
Using the fact that $\bar{f}$ is only a function of $\pi^0$ due to the term
$\delta(\pi_c\pi^c)$,
\begin{equation}
\Lie_{T \xi_1}\left[\bar{f}(x^{\nu},a p^{\mu}) \right]=T \frac{\partial}{\partial \eta}\Big|_{p}\bar{f}(x^{\nu},a p^{\mu}) + \frac{\partial \bar{f}}{\partial \pi^0}\pi^0 (T'+ n^i \dbi T)
\end{equation}
\begin{equation}
\left[\Tr\left(R^{(1)0}_{0,X}\right)+S^{(1)0}_{0,X}\right]\pi^0 \dsd{\bar{f}}{\pi^0}=-\frac{\partial
  \bar{f}}{\partial \pi^0}\pi^0(T'+ \HH T).
\end{equation}
Note that there is no term involving
$\left[\Tr\left(R^{(1)0}_{i,X}\right)+S^{(1)0}_{i,X}\right]\pi^i
\dsd{\bar{f}}{\pi^0}$ thanks to the prescription in the choice of the tetrad
in \S~\ref{sec:choicetetrad}.

We then express the derivatives as
\begin{equation}\label{derptopi}
\dsd{\bar{f}(x^{\nu},a p^{\mu})}{\eta}\Big|_{p}= \dsd{\bar{f}}{\eta}\Big|_{\pi} + \dsd{\bar{f}}{\pi^0} \HH \pi^0.
\end{equation}

Putting all the pieces together, we finally get that

\begin{eqnarray}\label{Tf1}
\Tr\left[ \delta(\pi_c\pi^c)\delta^{(1)}_X f\right]&=&\delta(\pi^c\pi_c)\left(\frac{\partial \bar{f}}{\partial
  \pi^0} \pi^0 n^i \dbi T + T\dsd{\bar{f}}{\eta}\Big|_{\pi}\right) \nonumber\\
&=&\delta(\pi^c\pi_c)\frac{\partial \bar{f}}{\partial
  \pi^0} \pi^0 (\HH T +n^i \dbi T), 
\end{eqnarray}
where in the last step we have made use of the background Boltzmann equation (\ref{Boltzmann0}).

It can be checked that by considering $f$ as a function
of $\sqrt{\pi^i \pi_i}$ instead of $\pi^0$, as allowed by the factor $\delta(\pi^c\pi_c)$,
we recover the same result as performed in Ref.~\cite{Ruth}. However this is
slightly more intricate, as it now apparently depends on the three variables
$\pi^i$ which are in fact not independent at the background level. 

Although the mathematical framework can seem to be heavy, we did not need to define an extension of the distribution function outside
the mass shell nor a gauge transformation field parallel to the mass shell as in Ref.~\cite{Ruth}.
We first have built the distribution function using the tetrad field (it is a
function of $\pi^a$ and not an express function of $p^{\mu}$). Then, as
explained in \S \ref{mass_shell}, the normalization condition
(\ref{deftetrad}), when expressed at each order in Eqs.~(\ref{Constraints}),
ensures that it remains on the mass shell during a gauge transformation that
we perform using the rules derived for tensors.

\subsection{First-order gauge-invariant distribution function for radiation}

Now that transformation properties of the first-order distribution
function are known, we can use the results of \S~\ref{sec1} to define a gauge-invariant distribution function by

\begin{eqnarray}
\hat{f}^{(1)}  &\equiv& \delta^{(1)}_{NG} f = \delta^{(1)}_X f + \Tr_{\xi^{(1)}_{\rightarrow NG}} \left(\delta^{(1)}_X f\right) \nonumber\\
&=& \delta^{(1)}_X f+\frac{\partial \bar{f}}{\partial  \pi^0} \pi^0 \left[\HH \left(B^{(1)}-E^{(1)'}\right) +n^i \dbi \left(B^{(1)}-E^{(1)'}\right)\right]. 
\end{eqnarray}

As for tensorial quantities, we can choose for instance
$\xi^{(1)}_{\rightarrow FG}$ in the above expression, in order to define an
other gauge-invariant distribution function. Its expression is given by

\begin{eqnarray}
\tilde{f}^{(1)}  &\equiv& \delta^{(1)}_{FG} f = \delta^{(1)}_X f + \Tr_{\xi^{(1)}_{\rightarrow FG}} \left(\delta^{(1)}_X f\right) \nonumber\\
&=& \delta^{(1)}_X f + \frac{\partial \bar{f}}{\partial  \pi^0} \pi^0 \left[\Psi^{(1)} + \frac{n^i \dbi\Psi^{(1)}}{\HH}\right]. 
\end{eqnarray}

These two first-order gauge-invariant distribution functions are related by

\begin{equation}
\tilde{f}^{(1)}-\hat{f}^{(1)}=\frac{\partial \bar{f}}{\partial  \pi^0} \pi^0 \left[\hat{\Psi}^{(1)} + \frac{n^i \dbi\hat{\Psi}^{(1)}}{\HH}\right].
\end{equation}

It is worth remarking that in the previous literature \cite{Ruth}, another
gauge-invariant distribution is defined, namely 
\begin{eqnarray}
{F}^{(1)}  &\equiv &\delta^{(1)}_X f+\frac{\partial \bar{f}}{\partial  \pi^0} \pi^0 \left[\Psi^{(1)} +n^i \dbi \left(B^{(1)}-E^{(1)'}\right)\right]\nonumber\\
&=&\hat{f}^{(1)}+\frac{\partial \bar{f}}{\partial  \pi^0} \pi^0 \hat{\Psi}^{(1)} 
\end{eqnarray}
Though it cannot be interpreted as the perturbation of the distribution function in a given gauge since it mixes $\xi_{\rightarrow NG}$ and $\xi_{\rightarrow FG}$, this is a better variable to highlight the conformal invariance of the photons propagation and to compare with the null cone integration method \cite{Ruth95}.   

This first-order analysis illustrates the power of this formalism which can be generalized to higher orders in perturbations.

\subsection{First-order collisionless Boltzmann equation for radiation}

Integrating the gauge-invariant distribution function of radiation over $\pi^0$, we define
the gauge-invariant brightness, which is the energy perturbation per unit
solid angle in a given direction

\begin{equation}
\hat{{\cal I}}^{(1)}(x^{\mu},n^i)\equiv 4 \pi\int \hat{f}^{(1)}(x^{\mu},\pi^0,n^i) (\pi^0)^3 \dd \pi^0.
\end{equation}

We choose the normalization of the background distribution function such that
the background brightness reduces to the energy density (see \S~\ref{sec:fluide}
for the fluid approximation)

\begin{equation}
\bar{{\cal I}}(\eta)\equiv 4 \pi\int \bar{f}(\eta,\pi^0)  (\pi^0)^3 \dd \pi^0= \bar{\rho}. 
\end{equation}
We can associate gauge-invariant symmetric trace-free moments, $\hat{{\cal F}}_{i_1...i_n}$, to this
brightness by using the decomposition

\begin{equation}\label{defGImoments1}
\hat{{\cal I}}^{(1)}(x^{\mu},n^i) \equiv \sum_p \hat{{\cal F}}^{(1)}_{i_1..i_p}(x^{\mu})n^{i_1}..n^{i_p}.
\end{equation}
With these definitions, the integral $\int
\left(\pi^0 \right)^3 \dd \pi^0$ on the first-order Boltzmann equation leads to the evolution equation for $\hat{{\cal I}}^{(1)}$ \cite{Hu}
\begin{equation}\label{bright1}
\left(\dsd{}{\eta}+ n^i \dbi\right)\frac{\hat{{\cal I}}^{(1)}}{4}+ \HH \hat{{\cal I}}^{(1)}  +  \left( n^i
\dbi \hat{\Phi}^{(1)} - \hat{\Psi}^{(1)'} \right)\bar{{\cal I}} =0, 
\end{equation}
where we have ignored the tensor terms for simplicity.
Similarly, a gauge-invariant brightness $\tilde{{\cal I}}^{(1)}$ associated with $\tilde{f}^{(1)}$, and a gauge-invariant brightness ${M}^{(1)}$ \cite{Ruth} associated with ${\cal F}^{(1)}$ can be defined. They are related to $\hat{\cal I}^{(1)}$ by

\begin{eqnarray}
\tilde{{\cal I}}^{(1)} &=& \hat{\cal I}^{(1)}-4 \bar{\cal I}\left( \hat{\Psi}^{(1)}+\frac{n^i \dbi \hat{\Psi}^{(1)}}{\HH}\right)\nonumber\\
{M}^{(1)} &=& \hat{\cal I}^{(1)}-4 \bar{\cal I} \hat{\Psi}^{(1)}.
\end{eqnarray}


\subsection{Gauge transformation at second order}

At second order, the general gauge transformation of the distribution function
(\ref{Tcompacte}) for $(\xi)=(\xi_1,\xi_2)$, $(T\xi)=(T\xi_1,T\xi_2)$ is given in details in \ref{app_Tf2}. After simplifications, it reads 

\begin{eqnarray}\label{Tf2}
&&\Tr\left( \delta^{(2)}_X f\right) =\dsd{\bar{f}}{\eta}(T^{(2)}+T T'+ \dbi T \dhi L) \nonumber\\
&&+\dsd{\bar{f}}{\pi^0}\pi^0 \Big\{ n^i \dbi T^{(2)} -2 n^j \left[
  \left( \dbi \dbj  E + E_{ij} + \dbi \dbj L\right) \dhi T - \Psi \dbj T\right]  \nonumber\\
&& \qquad \qquad  + \dbi T \dhi T+ (T n^i \dbi  T)'+ n^i\dbi \left( \dhj L \dbj T\right) + 2 \Phi n^i \dbi T\Big\}\nonumber\\
&& + \frac{\partial^2 \bar{f}}{\partial
  \left(\pi^0\right)^2}\left(\pi^0\right)^2 \left( n^i \dbi T n^j
  \dbj T\right) +2\frac{\partial^2 \bar{f}}{\partial \eta  \partial \pi^0}T n^i \dbi T  + \frac{\partial^2 \bar{f}}{\partial
  \eta^2}T^2 \nonumber\\
&& + 2 \dsd{\delta_X^{(1)} f}{\pi^0} \pi^0 n^j \dbj T+ 2 \dsd{\delta_X^{(1)}
  f}{\pi^i} \pi^0 \dhi T+ 2 \dhi L \dbi\delta_X^{(1)} f + 2 T \dsd{\delta_X^{(1)} f}{\eta}. 
\end{eqnarray}

This is a cornerstone expression in our study of the second-order distribution
function. As for the fluid quantities, knowing the transformation rules under
a second-order gauge change is enough to define a second-order gauge invariant
distribution function which is required to write the second-order Boltzmann
equation only in terms of gauge-invariant variables. As for tensors, several
gauge-invariant distribution function can be defined, and this relation is
also required to express how the different gauge-invariant distribution
functions are related.

\subsection{Second-order gauge-invariant distribution function for radiation}

Again, we can use the results of \S~\ref{defGIV} to define a gauge-invariant distribution function as

\begin{equation}
\hat{f}^{(2)} \equiv \delta^{(2)}_{NG}f  = \delta^{(2)}_X f + \Tr_{\left(\xi^{(1)}_{\rightarrow NG},\,\xi^{(2)}_{\rightarrow NG}\right)} \left[\delta^{(2)}_X f \right]. 
\end{equation}

As for tensorial quantities, we can choose for instance $\left(\xi^{(1)}_{\rightarrow FG},\,\xi^{(2)}_{\rightarrow FG}\right)$, in order to build another second-order gauge-invariant distribution function. 

\begin{equation}
\tilde{f}^{(2)} \equiv \delta^{(2)}_{FG}f  = \delta^{(2)}_X f + \Tr_{\left(\xi^{(1)}_{\rightarrow FG},\,\xi^{(2)}_{\rightarrow FG}\right)} \left[\delta^{(2)}_X f \right]. 
\end{equation}

The difference between these two gauge-invariant distribution functions is
also gauge-invariant and is consequently expressed only in terms of gauge
invariant quantities. For the sake of completeness, we give the form of the
relation between these two gauge-invariant distribution functions,

\begin{eqnarray}
&&\tilde{f}^{(2)}-\hat{f}^{(2)}=\nonumber\\
&&\frac{1}{\HH^2}\dsd{\bar{f}}{\pi^0}\pi^0 \Bigg\{ n^k \dbk
\left[\HH \hat{\Psi}^{(2)}+\hat{\Psi}^{(1)}\hat{\Psi}^{(1)'}+ 2 \HH
  \hat{\Psi}^{(1)2}\right] + \dbi \hat{\Psi}^{(1)} \dhi \hat{\Psi}^{(1)}\nonumber\\
&& \qquad \qquad + n^k \dbk \left[-\frac{\Delta^{-1}}{2 \HH}\left(\Delta
      \hat{\Psi}^{(1)} \right)^2 + \frac{\Delta^{-1}}{2 \HH}\left(\dhi \dbj \hat{\Psi}^{(1)}\dhj
        \dbi \hat{\Psi}^{(1)} \right)\right]  \nonumber\\
&& \qquad \qquad -2 \HH n^j E_{ij} \dhi \hat{\Psi}^{(1)}  +
\left(\hat{\Psi}^{(1)} n^i \dbi  \hat{\Psi}^{(1)}\right)' + 2 \HH \left(\hat{\Phi}^{(1)}+\hat{\Psi}^{(1)} \right) n^i \dbi \hat{\Psi}^{(1)}\Bigg\}\nonumber\\
&& + \frac{1}{\HH^2}\frac{\partial^2 \bar{f}}{\partial \left(\pi^0\right)^2
}\left(\pi^0\right)^2 \left[ n^i \dbi \hat{\Psi}^{(1)} n^j
  \dbj \hat{\Psi}^{(1)}\right] +\frac{2}{\HH^2}\frac{\partial^2
  \bar{f}}{\partial \eta  \partial \pi^0}\hat{\Psi}^{(1)} n^i \dbi
\hat{\Psi}^{(1)}  \nonumber\\
&& +\frac{1}{\HH^2}\dsd{\bar{f}}{\eta}\Big[\HH
  \hat{\Psi}^{(2)}+\hat{\Psi}^{(1)}\hat{\Psi}^{(1)'}+ 2 \HH \hat{\Psi}^{(1)2}+\HH
  \hat{\Psi}^{(1)}\left(\frac{\hat{\Psi}^{(1)}}{\HH} \right)' \nonumber\\
&& \qquad \qquad -\frac{\Delta^{-1}}{2 \HH}\left(\Delta
      \hat{\Psi}^{(1)} \right)^2 + \frac{\Delta^{-1}}{2 \HH}\left(\dhi \dbj \hat{\Psi}^{(1)}\dhj
        \dbi \hat{\Psi}^{(1)} \right)\Big] \nonumber\\
&& + \frac{2}{\HH} \dsd{\hat{f}^{(1)}}{\pi^0} \pi^0 n^j \dbj \hat{\Psi}^{(1)}+ \frac{2}{\HH} \dsd{\hat{f}^{(1)}}{\pi^i} \pi^0 \dhi \hat{\Psi}^{(1)} + 2 \frac{\hat{\Psi}^{(1)}}{\HH} \dsd{\hat{f}^{(1)}}{\eta}+ \frac{1}{\HH^2}\frac{\partial^2 \bar{f}}{\partial
  \eta^2}\left(\hat{\Psi}^{(1)}\right)^2  . 
\end{eqnarray}
This clearly demonstrates the power of our formalism since, contrary to the
first order, this relation cannot be guessed intuitively. Note also that this
is non-local as it is generally the case for second-order gauge-invariant quantities. 

\subsection{The second-order gauge-invariant collisionless Boltzmann equation
  for radiation}

Similarly to the first order case, we define the second-order brightness as

\begin{equation}
\hat{{\cal I}}^{(2)}(x^{\mu},n^i)\equiv 4 \pi\int \hat{f}^{(2)}(x^{\mu},\pi^0,n^i) (\pi^0)^3 \dd \pi^0.
\end{equation}
We also define the second-order gauge-invariant moments associated to this gauge
invariant brightness by the second-order version of Eq.~(\ref{defGImoments1}).  
The derivation of the collisionless Boltzmann equation in the Newtonian
gauge is detailed in Ref.~\cite{BartoloBoltzmann2_I,BartoloBoltzmann2_II}. Once the integral $\int
\left(\pi^0 \right)^3 \dd \pi^0$ performed, it
leads to an evolution equation for the brightness. As this is a scalar equation, it is gauge
invariant and it can be expressed only in terms of the gauge invariant
quantities that we have defined and which reduce to the perturbation variables in the
Newtonian gauge. Explicitly, it reads

\begin{eqnarray}\label{bright2}
&&\left(\dsd{}{\eta}+ n^i \dbi\right)\frac{\hat{{\cal I}}^{(2)}}{4}+ \HH \hat{{\cal I}}^{(2)}  + \bar{{\cal I}} n^i
\dbi \hat{\Phi}^{(2)}+ 2 \bar{{\cal I}}\left(\Psi^{(1)}-\Phi^{(1)}
\right)n^i \dbi \Phi^{(1)}\nonumber\\
&&  + \frac{1}{2}\left[\dbj\left(\hat{\Phi}^{(1)}+\hat{\Psi}^{(1)}\right)n^i
n^j-\dhi\left(\hat{\Phi}^{(1)}+\hat{\Psi}^{(1)}\right)
\right]\dsd{\hat{{\cal I}}^{(1)}}{n^i}\nonumber\\
&& -2 \hat{{\cal I}}^{(1)} \left( \hat{\Psi}^{(1)'} - n^j\dbj
  \hat{\Phi}^{(1)} \right) - \bar{{\cal I}}\left(\hat{\Psi}^{(2)'}+4\hat{\Psi}^{(1)}\hat{\Psi}^{(1)'} \right)\nonumber\\
&& +\frac{1}{2}\left(\hat{\Phi}^{(1)}+\hat{\Psi}^{(1)}\right)n^i \dbi \hat{{\cal I}}^{(1)} =0. 
\end{eqnarray}
Up to this stage, we agree with the expressions of Ref.~\cite{BartoloBoltzmann2_I,BartoloBoltzmann2_II}.

\section{Fluid approximation}\label{sec:fluide}

If we want to recover the transformation rule and the gauge-invariant variables for the energy density, the
pressure and the velocity of radiation, we need to define a stress-energy tensor from the
distribution function of radiation. We already know from special relativity
how to define such a tensor. We generalize it by using the local Minkowskian frame

\begin{eqnarray}\label{defTab}
T^{ab}(x^{\mu})&=& \int \dd \pi^0 \dd^3 \pi^i \delta(\pi^c \pi_c)
f(x^{\mu},\pi^d)\pi^a \pi^b \nonumber\\
&=& \int (\pi^0)^3 f(x^{\mu},\pi^d)n^a n^b \dd \pi^0 \dd^2 n^i,
\end{eqnarray}
where $n^a\equiv n^i = \pi^i / \pi^0$, if $a=1,2,3$ and $n^a=1$ if $a=0$.
In order to evaluate the stress energy tensor, we have performed one of the integrals
which removes the Dirac contribution $\delta(\pi^a \pi_a)$
\begin{equation}
\int \delta(\pi^a \pi_a) G(x,\pi^a)\dd \pi^0 \dd^3 \pi^i= \int G(x,\pi^0,
n^i) \pi^0 \dd \pi^0 \dd^2 n^i. 
\end{equation}
Several useful relations for handling integrals of the background distribution
function are reported in \ref{app_int_utiles}. 
If we are dealing with several species, we can still define a stress-energy tensor for each component, as long as we are dealing with weakly interacting gases. This is the standard kinetic approach in which the interaction between two species is encoded in the collision term of the Boltzmann equation \cite{Ehlers-71}.

We define $\rho$, $P$, the velocity $U^a$ and the anisotropic stress $\Pi^{ab}$, 
\begin{equation}\label{decT}
T^{ab}= \rho U^a U^b + P\perp^{ab} + \Pi^{ab}, 
\end{equation}
with $\perp^{ab}\equiv\eta^{ab}+U^aU^b$, and the properties $U^a U_a
=-1,\,\,\Pi^{ab}\perp_{ab}=0$, $U_a \Pi^{ab}=0$.
However, fluid quantities are usually expressed using the canonical basis
associated with coordinates $\partial_{\mu}$ and not the tetrad field. We thus
define $u^{\mu}=U^a e_a^{\mu}$ as the coordinates of the velocity in this
canonical basis, and we decompose it as in Eq.~(\ref{defvelocity}). Similarly,
we define the anisotropic stress expressed in the canonical basis by $\pi^{\mu
  \nu} = e^{\mu}_{a} e^{\nu}_{b} \Pi^{ab} $. Some confusion can arise from the
fact that physicists often design a vector by its coordinates. With this symbolic convention, $U^a$ and $u^{\mu}$ are
mathematically the same vector, but expressed in different basis since $U^a
e_a = u^{\mu} \partial_{\mu}$. The relations
between $U^a$ and $u^{\mu}$ up to second order are 
\begin{eqnarray}\label{Utou0}
\bar{U}^0 &=& a \bar{u}^{0} = 1\nonumber\\
\bar{U}^i &=& a \bar{u}^{i} = 0,
\end{eqnarray}
and
\begin{eqnarray}\label{Utou1}
\delta_X^{(1)} U^0 &=& 0\nonumber\\
\delta_X^{(1)} U^i &=& \dhi\left(v^{(1)}+B^{(1)}\right),
\end{eqnarray}
\begin{eqnarray}\label{Utou2}
\delta_X^{(2)} U^0 &=& \dbi(v + B)\dhi(v+B)\nonumber\\
\delta_X^{(2)} U^{i} &=& \dhi(v^{(2)}+B^{(2)}) - 2\Phi\dhi B  + 2\Psi \dhi \left(B-v\right)   \nonumber\\
&& +2 \dhj\left(v-B\right) \left(\dhi \dbj E + E^i_{\,\,j}\right).
\end{eqnarray}
Similarly the relations between the spatial components of $\pi^{\mu \nu}$ and $\Pi^{ab}$ are
\begin{eqnarray}\label{pitoPi}
\delta_X^{(1)} \pi^{ij} & = & \frac{1}{a^2}\delta_X^{(1)} \Pi^{ij}\nonumber\\
\delta_X^{(2)} \pi^{ij} & = & \frac{1}{a^2} \Big[ \delta_X^{(2)}
  \Pi^{ij}+2\delta_X^{(1)} \Pi^{ik}\left(  \Psi^{(1)}\delta_{k}^{j} - \dbk
    \dhj E^{(1)} -E^{(1)j}_{k}  \right)\nonumber\\
&& \qquad \qquad +2\delta_X^{(1)} \Pi^{jk}\left(  \Psi^{(1)}\delta_{k}^{i} - \dbk
    \dhi E^{(1)} -E^{(1)i}_{k}  \right) \Big].
\end{eqnarray}

The fluid quantities can be extracted from Eq.~(\ref{decT}) as follows
\begin{eqnarray}
&&\rho=T^{ab}U_aU_b,\label{defrho}\\
&&3P=T^{ab}\perp_{ab},\label{defP}\\
&&\Pi_{ab}=T^{cd}\left(\perp_{ca}\perp_{db}-\frac{1}{3}\perp_{cd}\perp_{ab}\right),\label{defPi}\\
&&(\rho+P)U^0U^{i}=T^{0i}.\label{defv}  
\end{eqnarray}
It is easy to see that the factor $\delta\left(\pi_a \pi^a\right)$ in the
integral of the definition (\ref{defTab}) of the stress energy tensor implies that $P=\rho/3$. 

The system of definitions~(\ref{defrho}-\ref{defv}) determines the fluid quantities. Indeed, these
quantities can now be calculated iteratively at any order once
Eq.~(\ref{defTab}) is pulled back to the background space-time. Since
$\bar{U}^0=1$ and $\bar{U}^{i}=0$, $\bar{\rho}$ $\bar{P}$ and $\bar{\Pi}^{ab}$ are given by

\begin{equation}
\bar{\rho} = 3 \bar{P} = \bar{T}^{00} \bar{U}_0 \bar{U}_0,\quad\bar{\Pi}^{ab}=0,
\end{equation}
as expected from the background symmetries for a fluid of radiation.
Then, since $U^0 = \sqrt{U^{i} U_{i}+1}$, and using Eq.~(\ref{defv}), we can
determine the first-order expression of the velocity
\begin{eqnarray}\label{U1}
\delta_X^{(1)}U^{0} &=& 0\nonumber\\
\delta_X^{(1)}U^{i} &=& \frac{3}{4 \bar{\rho}} \delta_X^{(1)} T^{0i}.
\end{eqnarray}
 Repeating this procedure, we obtain from Eqs.~(\ref{defrho}-\ref{defv})
\begin{eqnarray}\label{rho1}
\delta_X^{(1)} \rho &=& 3 \delta_X^{(1)} P= \delta_X^{(1)} {T}^{00} \bar{U}_0
\bar{U}_0 \nonumber\\
\delta_X^{(1)} \Pi^{ij} &=& \delta_X^{(1)} T^{ij}-\frac{\delta^{ij}}{3}\delta_X^{(1)} T^{k}_{\,\,k}\,,\label{pi1}
\end{eqnarray}
and the condition $U_a \Pi^{ab}=0$ implies
\begin{eqnarray}\label{Pi0i}
\delta_X^{(1)} \Pi^{i0} &=& \delta_X^{(1)} \Pi^{00}=0\nonumber\\
\delta_X^{(2)} \Pi^{00} &=& 0\nonumber\\
\delta_X^{(2)} \Pi^{0i} &=& 2 \delta_X^{(1)} \Pi^{ij}\delta_X^{(1)}U_{j}.   
\end{eqnarray}
Again, using Eq.~(\ref{defv}), we determine the second-order perturbation of the velocity 

\begin{eqnarray}
\delta_X^{(2)}U^{0} &=& \delta_X^{(1)}U^{i} \,\delta_X^{(1)}U_{i} \label{U20}\\
\delta_X^{(2)}U^{i} &=& \frac{3}{4 \bar{\rho}} \left(\delta_X^{(2)} T^{0i} - \delta_X^{(2)} \Pi^{0j}\right) -
2\frac{\delta_X^{(1)} \rho}{\bar{\rho}} \delta_X^{(1)} U^{i}.\label{U2i}
\end{eqnarray}
Iterating, we obtain from Eqs~(\ref{defrho}-\ref{defv})
\begin{eqnarray}
\delta_X^{(2)} \rho &=& 3 \delta_X^{(2)} P= \delta_X^{(2)} {T}^{00} \bar{U}_0
\bar{U}_0 +2 \bar{T}^{00} \bar{U}_0 \delta_X^{(2)} U_0 \label{rho2}\\
\delta_X^{(2)} \Pi^{ij} &=& \delta_X^{(2)}
T^{ij}-\frac{\delta^{ij}}{3}\delta_X^{(2)}
T^{k}_{\,\,k}\nonumber\\
&& \,-\frac{8}{3}\bar{\rho}\left(\delta_X^{(1)}U^i \delta_X^{(1)} U^j
  - \frac{\delta^{ij}}{3}\delta_X^{(1)}U^k\,\delta_X^{(1)}U_k \right).\nonumber
\end{eqnarray}

This shows that, by iterating this procedure, the fluid quantities can be determined up to order
  $n$ if $f$, that is $T^{ab}$, is known up to order $n$. This means that, by knowing the transformation rule of $f$ under a gauge transformation, we can deduce the transformation rules of the
  fluid quantities built out of it ($\rho,\,P,\,U^a,\,\Pi^{ab}$). Eventually, we are
  interested in their expressions in the canonical basis in order to compare
  with the results of \S~\ref{sec1}, and we need to use
  Eqs.~(\ref{Utou0}-\ref{Utou2}) and Eqs.~(\ref{pitoPi}).

\subsection{First-order fluid quantities transformation}

At first order, from the relations (\ref{rho1}) and (\ref{Utou1}), and  the
transformation rule for $f$, Eq.~(\ref{Tf1}), we deduce after some algebra,
that $\delta^{(1)}_X \rho$ transforms as in Eq.~(\ref{Tfluide1}). Similarly, from Eq.~(\ref{U1}), the relation (\ref{Utou1}), and the transformation
rule for $f$, Eq.~(\ref{Tf1}), we deduce that $v^{(1)}$ transforms as in Eq.~(\ref{Tfluide1}). By the same method, we recover easily that $\delta^{(1)}
\pi^{ij}$ is gauge invariant.

\subsection{First-order fluid equations}

In order to recover the gauge-invariant conservation equation and the Euler
equation of the fluid approximation at first order, we define the first-order gauge invariant
stress-energy tensor by

\begin{equation}\label{defGITab1}
\hat{T}^{ab(1)}(x^{\mu}) \equiv \int (\pi^0)^3 \hat{f}^{(1)}(x^{\mu},\pi^c)n^a
n^b \dd \pi^0 \dd^2 \Omega= \int \hat{\cal I}^{(1)} n^a n^b \frac{\dd^2 \Omega}{4 \pi},
\end{equation}
and its associated first-order gauge-invariant fluid quantities,
$\hat{\rho}^{(1)}$, $\hat{P}^{(1)}$, $\hat{v}^{(1)}$ and $\hat{\pi}^{ij(1)}$,
built from the same types of relation as in the set of
Eqs.~(\ref{defrho}-\ref{defv}) and expressed in the canonical basis with
Eqs·~(\ref{Utou1}) and (\ref{pitoPi}). Because of the comparison
performed in the previous section, these quantities match those defined in
Eq.~(\ref{deffluideGI1}), and this justifies the fact that we use the same notation.
We need the useful relations between the first moments and the fluid quantities

\begin{equation}
\hat{\cal F}^{(1)} = \int \hat{{\cal I}}^{(1)} \frac{\dd \Omega}{4 \pi} = \delta^{(1)}\hat{\rho}, 
\end{equation}
\begin{equation}
\hat{\cal F}^{i(1)} = \int \hat{{\cal I}}^{(1)} n^i \frac{\dd \Omega}{4 \pi} =
\frac{4}{3}\bar{\rho}\dhi \hat{v}^{(1)},
\end{equation}
\begin{equation}
\hat{\cal F}^{ij(1)} = \int \hat{{\cal I}}^{(1)} \left(n^in^j
  -\frac{\delta^{ij}}{3} \right)\frac{\dd \Omega}{4 \pi} = \hat{\Pi}^{ij(1)}.
\end{equation}
Performing $\int \dd \Omega $ on the brightness evolution equation
(\ref{bright1}), we recover the first-order conservation equation. However,
performing  $\int n^i \dd \Omega $, we recover the first
order Euler equation as expressed in \ref{app_fluid_equations}, {\it only if} we
neglect the first-order anisotropic pressure. This comes from the fact that the statistical
description of radiation leads to an infinite hierarchy of equations coupling
moments of order $p-1$, $p$ and $p+1$ \cite{Hutotal}, whereas the fluid
description keeps only the equations involving the monopole and the dipole.

\subsection{Second-order fluid quantities transformation}

In order to establish the second-order comparison with the fluid description, we
need to know how to perform an integral involving $\delta_X^{(1)} f$, for
instance on  $ 2 \dsd{\delta_X^{(1)} f}{\pi^0} \pi^0 n^j \dbj T+ 2 \dsd{\delta_X^{(1)}
  f}{\pi^i} \pi^0 \dhi T$. We will thus make use of the multipolar expansion
\begin{equation}
\delta^{(1)}_X f = \bar{f} \frac{\delta^{(1)}_X \rho}{\bar{\rho}} + 4 \bar{f}\dbi\left(v^{(1)} + B^{(1)}\right) n^i + \frac{15 \bar{f}}{2 \bar{\rho}} \delta^{(1)}_X \Pi_{ij}n^i n^j+...
\end{equation}  
from which it can be checked that we recover the correct fluid quantities when used to compute $\delta^{(1)}_X T^{ab}$ in Eq.~(\ref{defTab}).

Using the same method as for the first order, with the relations (\ref{rho2}) and
(\ref{Utou2}), and the transformation rule for the second-order distribution
function, Eq.~(\ref{Tf2}), we deduce that $\delta^{(2)} \rho$ transforms as in Eq.~(\ref{transfo_order_2}). 
Additionally, from the relations (\ref{U20}), (\ref{U2i}), (\ref{Utou2}) and (\ref{Pi0i}), we deduce that $v^{(2)}$ transforms as in Eq.~(\ref{transfo_order_2}). 

We also notice that from the definition (\ref{rho2}), the
relations (\ref{U20}) (\ref{U2i}), and the transformation rule for $f$, Eq.~(\ref{Tf2}), we
deduce that $\delta^{(2)} \Pi^{ij}$ transforms according to
\begin{equation}
\delta^{(2)} \Pi^{ij} \rightarrow \delta^{(2)} \Pi^{ij} + 2 T \left(\delta^{(1)}
\Pi^{ij}\right)' + 2 \dhk L \dbk \left(\delta^{(1)} \Pi^{ij}\right). 
\end{equation}
When expressed in the canonical basis ($\pi^{\mu \nu} \equiv e^{\mu}_{a}
e^{\nu}_{b} \Pi^{ab}$), we recover exactly the transformation
rule of the anisotropic stress given in Eq.~(\ref{transfo_order_2}).

This is one of the major results of this paper. We recover the perfect fluid
transformation rules for the energy density, the pressure, the velocity and the anisotropic stress given in Ref.~\cite{Bruni-1998} up to second order, when
starting from the statistical description.

\subsection{Second-order fluid equations}

In order to recover the gauge-invariant conservation equation and the Euler
equation of the fluid approximation at the second order, we follow the same
procedure as for the first order case. We thus define the second-order gauge invariant
stress-energy tensor by

\begin{equation}\label{defGITab2}
\hat{T}^{ab(2)}(x^{\mu}) \equiv \int (\pi^0)^3 \hat{f}^{(2)}(x^{\mu},\pi)n^a n^b \dd \pi^0 \dd^2 \Omega= \int \hat{\cal I}^{(2)} n^a n^b \frac{\dd^2 \Omega}{4 \pi},
\end{equation}
and its associated second-order gauge-invariant fluid quantities, $\hat{\rho}^{(2)}$
$\hat{P}^{(2)}$ $\hat{v}^{(2)}$ and $\hat{\pi}^{ij(2)}$, built from the same
types of relations as in the set of Eqs.~(\ref{defrho}-\ref{defv}) and
expressed in the canonical basis with Eqs·~(\ref{Utou2}) and (\ref{pitoPi}). Because of the comparison
performed in the previous section, these quantities match those defined in Eq.~(\ref{deffluideGI2}), thus justifying the fact
that we use the same notation.

In order to recover the conservation and Euler equations of the fluid approximation we
perform the integral $\int \frac{\dd \Omega}{4 \pi} $ and $\int \frac{\dd
  \Omega}{4 \pi} n^i$ on this equation. However, at the second order this has
to be done with care since the link between the second-order gauge-invariant brightness and the second-order fluid quantities is given by

\begin{equation}
\hat{\cal F}^{(2)}  = \int \hat{{\cal I}}^{(2)} \frac{\dd \Omega}{4 \pi} =
\delta^{(2)}\hat{\rho} + \frac{8}{3} \bar{\rho}\dbi\hat{v}^{(1)}
\dhi\hat{v}^{(1)},
\end{equation}
\begin{equation}
\hat{\cal F}^{i(2)} =\int \hat{{\cal I}}^{(2)} n^i \frac{\dd \Omega}{4 \pi} =
\frac{4}{3}\bar{\rho}\left(\dhi \hat{v}^{(2)} -2 \hat{\Psi}^{(1)} \dhi \hat{v}^{(1)}\right) + \frac{8}{3} \delta^{(1)} \hat{\rho} \dhi
  \hat{v}^{(1)},
\end{equation}
\begin{eqnarray}
\hat{\cal F}^{ij(2)} &=& \int \hat{{\cal I}}^{(2)} \left(n^in^j
  -\frac{\delta^{ij}}{3} \right)\frac{\dd \Omega}{4 \pi} \nonumber\\
& = &\hat{\Pi}^{ij(2)}+\frac{8}{3}\bar{\rho}\left[\dbi \hat{v}^{(1)} \dbj \hat{v}^{(1)} - \frac{\delta_{ij}}{3} \left( \dbk \hat{v}^{(1)} \dhk \hat{v}^{(1)} \right) \right].\label{ItoPi2}
\end{eqnarray}
This clearly differs from the expressions (5.10) and (6.33) of Ref.~\cite{BartoloBoltzmann2_I}
where the term quadratic in $v$ in $\hat{\cal F}^{(2)}$, the term quadratic in
$\Psi$ and $v$ in $\hat{\cal F}^{i(2)}$ are not there. The difference in the
energy density perturbation as extracted from $\hat{\cal
  F}^{(2)}$, comes from the fact that the fractional energy density $\Delta^{(2)}$ for the radiation is defined as seen by the observer of
velocity $ e^0_{\mu} \sim (\dd \eta)_{\mu}$ whereas we define it in the fluid
frame. The fractional energy density that they define is related to
our quantities by $\bar{\rho} \Delta^{(2)} = \delta_{NG}^{(2)} {T}^{00} \bar{U}_0
\bar{U}_0$. The difference in the expressions for the fractional energy
density can be traced using Eqs.(\ref{rho2}) with Eq.(\ref{U20}). However, this
is only a matter of definition and it is consistent with Eq.(7.2) of
Ref.~\cite{BartoloBoltzmann2_I}. Implicitly the authors of Ref.~\cite{BartoloBoltzmann2_I}
do also use a tetrad basis in their section 3 in order to identify coordinates of
the tangent space between the background and the perturbed space-time, in the
same way as explained below Eq.(\ref{decmultipoles}). Their $p$ is equal to our $\pi^0$ and the
unit vectors $n^i$ match when restricting to the Newtonian gauge. The
equations (3.6) and (3.7) of Ref.~\cite{BartoloBoltzmann2_I} are equivalent to
Eq.(\ref{ptopi}) when expressed in the newtonian gauge with the use of
Eqs.(\ref{defRS}), Eqs.(\ref{B1}) and Eqs.(\ref{B3}). As for the difference in the velocity
perturbation as defined from $\hat{\cal F}^{i(2)}$, it comes from the fact
that their definition for $v_{\gamma}^{i(2)}$has to be interpreted in the
tetrad basis, and therefore it matches $\delta_{NG}^{(2)}U^{i}$. However, the difference between the
tetrad basis and the canonical basis is not computed as in Eq.(\ref{U2i}), and
it explains the discrepancy. This can also be checked on the second-order
extraction of Eq.(7.3) in Ref.~\cite{BartoloBoltzmann2_I}. Indeed, there is
an the extra term quadratic in $\Psi$ and $v^i$ when compared to Eq.(2.15) of
Ref.~\cite{Bartolo2003}, as a trace of the difference between our perturbed
velocity, which matches the definition in the canonical basis usually given by
Eq.(\ref{defvelocity}) and Eq.(\ref{decomposition-ordre2}), and their perturbed velocity.
However, the equations involving $v^{i(2)}_{\gamma}$ in Refs.~\cite{BartoloBoltzmann2_I,BartoloBoltzmann2_II} such as Eq.(4.6) are consistent with this
difference, though the physical interpretation $v^{i(2)}_{\gamma}$ as being
the perturbed velocity of photons in the canonical basis is not correct.

The computation of a term like$\dsd{\hat{f}^{(1)}}{n^i}$, is easily performed using the multipolar expansion

\begin{equation}
\hat{f}^{(1)} = \bar{f} \frac{\delta^{(1)} \hat{\rho}}{\bar{\rho}} + 4 \bar{f}\dbi\hat{v}^{(1)} n^i + \frac{15 \bar{f}}{2 \bar{\rho}} \hat{\Pi}^{(1)}_{ij}n^i n^j+...
\end{equation}  
Applying this method, we recover the second-order conservation equation
detailed in \ref{app_fluid_equations}. As for the Euler equation, we recover it at second
order only if we neglect the anisotropic stress up to second order (beware
that the anisotropic stress is different from the second moment of the
distribution as it can be seen on Eq.~(\ref{ItoPi2})), and use the first-order Euler equation. 

This is also a major result of this paper. We recover the fluid gauge
invariant equations up to second order, only if we can neglect the anisotropic stress up to second order. It remains to be shown that this is extended up to any order, as we expect.

Let us also stress that in Ref.~\cite{BartoloBoltzmann2_II}, the term $\dsd{\hat{f}^{(1)}}{n^i}$ is evaluated using
$\dsd{\hat{f}^{(1)}}{n^i}=\dsd{\hat{f}}{x^j}\dsd{x^j}{n^i}$, in order to derive Eq.(4.1) and Eq.(4.6). However, this is not correct since $\hat{f}$ is a function of the independent variables
$\eta,x^i,\pi^0,n^i$. Even though they are related on a photon geodesic, they are independent in the analytic
expression of $\hat{f}$. Additionally this method is not fruitful because $\dsd{x^j}{n^i} \neq
\delta^{j}_{i}\left(\eta -\eta_i\right)$, since $n^i$ does not parameterize a photon geodesic. Consequently, the subsequent analytic expressions of
this reference solving the conservation and Euler equation are not correct
(for both radiation and cold dark matter) though the Boltzmann equation is
correct. This can also be seen directly from the fact that these equations do
not match fluid approximation equations of \ref{app_fluid_equations}. Once
corrected for this mistake. and taking into account the differences mentioned
before we can check that the collisionless part of the conservation and Euler equations (4.1) and (4.6) of Ref.~\cite{BartoloBoltzmann2_I} match our equations.

\subsection{Validity of the fluid approximation in the literature}

In this paper, we have considered so far the fluid approximation as a theoretical
framework in which we restrict the description of a species to its energy
density and its velocity. The computations involved for the distribution
function at second order were rather long, and it was used as a consistency check for the
gauge transformation rules and the collisionless Boltzmann equation. Since the fluid
approximation is built out of the kinetic theory, it was indeed expected that all the
conclusions made in this statistical description could find their fluid approximation counterpart.  

It is now necessary to determine under which conditions this can be done, that
is when the anisotropic stress can be neglected. This requires to work on the
physics of coupled species, baryons and photons, in the cosmological context.
The collision term as well as its physical implications have been studied in
Ref.~\cite{BartoloBoltzmann2_II} and it is very likely that the extraction of
its quadrupole in Eq.(4.18) is not affected by the previous
considerations. Indeed, in the tight coupling limit (which requires only the collision term)
for a system of photons and electrons highly coupled through the Compton diffusion, the authors of Ref.~\cite{BartoloBoltzmann2_II} find that the quadrupole satisfies
\begin{equation}
\hat{\cal F}^{ij(2)} \simeq\frac{8}{3}\bar{\rho}\left[\dhi \hat{v}^{(1)} \dhj \hat{v}^{(1)} - \frac{\delta^{ij}}{3} \left( \dbk \hat{v}^{(1)} \dhk \hat{v}^{(1)} \right) \right].  
\end{equation} 
This result is necessary to determine in which case the fluid approximation
can be used. Comparing it with Eq.~(\ref{ItoPi2}), we immediately see that the
physical interpretation of this result is that the second-order anisotropic stress of radiation $\hat{\Pi}^{ij(2)}$ is
suppressed in the tight coupling limit. As a consequence, the fluid approximation can be used in the tight
coupling limit also at second order in perturbations. 

\section{Conclusion}

In this article, we have performed a general investigation of the gauge
invariance of the distribution function. This allows us to recover very easily
the standard results at the first order and to extend them at the second order. We
derived the fluid approximation at first and second orders. This required to
carefully define the stress-energy tensor in the local Minkowskian frame. At
the second order, our results differ from the ones previously derived in the
literature \cite{BartoloBoltzmann2_I,BartoloBoltzmann2_II}. We have tackled down the origin of the differences and shown that it was lying in an incorrect identification between the tetrad and the
canonical basis. Our analysis, restricted to the collisionless case, puts the
second order Boltzmann equation, needed if we intend to study non-Gaussianities
in the CMB, on firm ground.

\section*{Acknowledgements}

I thank Jean-Philippe Uzan for drawing the topic to my attention and for his
endless comments on the manuscript. The second-order expansions were computed
using the tensor calculus package xAct \cite{JMM} and I thank Guillaume Faye and Jos\'e
Mart\'in Garc\'ia for their help on it. I thank Ruth
Durrer Nicola Bartolo Sabino Matarrese and Antonio Riotto for commenting on their works. Finally, I
thank Thiago dos Santos Pereira for his numerous remarks on the draft.

\appendix

\section{Sources terms in second order transformations}\label{app_sources}

The perturbation variables in the decomposition (\ref{metric}) are extracted
as follows
\begin{eqnarray}\label{extraction_metric}
\Phi &=& -\frac{1}{2 a^2} \delta g_{00},\\
\Psi &=& - \frac{1}{4 a^2}\left(\delta^{ij}- \Delta^{-1}\dhi \dhj\right) \delta g_{ij}, \nonumber\\
B &=& \frac{1}{a^2} \Delta^{-1} \dhi \delta g_{0i},\nonumber\\
E &=& \frac{1}{4 a^2} \left(\Delta \Delta \right)^{-1} \left(3\dhi \dhj- \Delta
  \delta^{ij}\right) \delta g_{ij} ,\nonumber\\
E_{pq} &=& \frac{1}{2 a^2}\left(\delta_{p}^{r}-\Delta^{-1}\partial_p \partial^r
\right)\left(\delta_{q}^{s}-\Delta^{-1}\partial_q \partial^s
\right)\left(\delta_{r}^{i} \delta_{s}^{j}-\frac{1}{3}\delta_{rs} \delta^{ij} \right) \delta g_{ij}.\nonumber
\end{eqnarray}
Using this method we can read the source terms defined in
Eq.~(\ref{transfo_order_2}), which are quadratic in the gauge change variables
$T,L$ and the perturbation variables $\Phi,\Psi,B,E,E_{ij}$, in Eq.~(\ref{transforule})

\begin{eqnarray}
S_{\Phi} &=& T\left(T''+ 5 \HH T' + (\HH'+2\HH^2)T +4 \HH \Phi+ 2\Phi'\right) \nonumber\\
&&+ T'\left(2T'+4 \Phi\right) + \dbi L \dhi\left( T' + \HH T + 2 \Phi\right) \nonumber\\
&&+ \dbi L' \dhi\left(T- 2 B - L'\right),\\
S_{\Psi} &=& - T \left(\HH T' + (\HH' + 2 \HH^2) T - 2 \Psi' -4 \HH \Psi\right) \nonumber\\
&&- \dbi\left(\HH T-2 \Psi\right)\dhi L\nonumber\\
&& - \frac{1}{2}\left(\delta^{ij}- \Delta^{-1}\dhi \dhj\right)\Bigg[\dbj\left(2 B+ L'-T\right)\dbi T\nonumber\\
&& + \dbi \dhk L \left( 2 \dbk \dbj L + 4\dbk \dbj E+4 E_{kj}+ (2 \HH T -4
\Psi) \delta_{kj}\right) \nonumber\\
&& + T \dbi \dbj \left(L'+2\HH L\right)\nonumber\\
&& + T \left(2 E'_{ij}+2 \dbi \dbj E' +4\HH E_{ij}+4 \HH \dbi \dbj E\right) \nonumber\\
&&+ \dhk L \dbk\left(\dbi \dbj L +2 E_{ij}+ 2 \dbi \dbj E\right) \Bigg]. 
\end{eqnarray}
$S_{\Psi}$ is slightly different from Ref.~\cite{Malikseul} and
Ref.~\cite{MalikWands} since, in these works, the extraction of metric
perturbation variables is not performed according to
Eq.~(\ref{extraction_metric}). However, this mistake does not matter for their
study that focused on the long wavelength limit.
\begin{eqnarray}
S_{B} &=&\Delta^{-1}\dhi \Big\{ T' \dbi(2 B + L' -T ) \nonumber\\
&&+ \dhj L' \left[2 \dbi \dbj L + 2 \left(\HH T-2\Psi\right) \delta_{ij} + 4 \left(E_{ij}+\dbi \dbj E \right)\right]  \nonumber\\
&&+ \dhj \dbi L \dbj\left(2 B + L'-T \right) + \dhj L \dbj \dbi \left(2B+L'-T\right) \nonumber\\
&&+ \dbi T (-4 \Phi -2 T' -2\HH T) + T \dbi(2 B' + L'' - T') \nonumber\\
&&+ 2 \HH T \dbi(2B+L'-T) \Big\},\\
S_{E}&=&(\Delta \Delta)^{-1} \left(\frac{3}{2}\dhi \dhj
  -\frac{1}{2}\Delta \delta^{ij}\right)\Big\{\dbj\left(2 B+ L'-T\right)\dbi T\nonumber\\
&& + \dbi \dhk L \left[ 2 \dbk \dbj L + 4\dbk \dbj E+4 E_{kj}+ (2 \HH T -4
\Psi) \delta_{kj}\right] \nonumber\\
&& + T \dbi \dbj (L'+2\HH L)\nonumber\\
&& + T \left(2 E'_{ij}+2 \dbi \dbj E' +4\HH E_{ij}+4 \HH \dbi \dbj E\right) \nonumber\\
&&+ \dhk L \dbk\left(\dbi \dbj L +2 E_{ij}+ 2 \dbi \dbj E\right)
\Big\},\\
{S_{E}}_{pq}&=&\left(\delta_{p}^{r}-\Delta^{-1}\partial_p \partial^r
\right)\left(\delta_{q}^{s}-\Delta^{-1}\partial_q \partial^s
\right)\left(\delta_{r}^{i} \delta_{s}^{j}-\frac{\delta_{rs}}{3} \delta^{ij} \right) \nonumber\\
&& \Big\{ \dbi \dhk L \left[ 2 \dbk \dbj L + 4\dbk \dbj E+4 E_{kj}+ (2 \HH T -4
\Psi) \delta_{kj}\right] \nonumber\\
&& + T \dbi \dbj (L'+2\HH L)+\dbj\left(2 B+ L'-T\right)\dbi T\nonumber\\
&& + T \left(2 E'_{ij}+2 \dbi \dbj E' +4\HH E_{ij}+4 \HH \dbi \dbj E\right) \nonumber\\
&&+ \dhk L \dbk\left(\dbi \dbj L +2 E_{ij}+ 2 \dbi \dbj E\right) \Big\},  
\end{eqnarray}

\begin{eqnarray}
S_{\rho}&=&T(\bar{\rho}''T + \bar{\rho}'T' + 2 \delta \rho ') + \dhi L \dbi (2 \delta \rho + \bar{\rho}' T),\\
S_{P}&=&T(\bar{P}''T + \bar{P}'T' + 2 \delta P ') + \dhi L \dbi (2 \delta P + \bar{P}' T),\\
S_{v}&=&\Delta^{-1}\dbi \Big[\HH T\dhi(L' - 2 v) + T\dhi(2 v' -L'') \nonumber\\
&&\qquad \quad  + L^j \dbj \dhi(2 v -L') + \dhi L'\left(\HH T + T' + 2\Phi\right) \nonumber\\
&& \qquad \quad + \dhj (L' -2 v)\dbj \dhi L  \Big].
\end{eqnarray}

\section{Transformation rules of the tetrad fields}\label{app_TRS}

$R_{ab}$ and $S_{ab}$ are defined in Eq.~(\ref{defRS}). The perturbation variables of the metric are defined in Eq.~(\ref{metric}).

\subsubsection{First order}

\begin{eqnarray}\label{B1}
R^{(1)}_{00,X} &=& -S^{(1)}_{00,X} =\Phi^{(1)}\\
R^{(1)}_{0 a_i,X} &=& -S^{(1)}_{0 a_i,X} =- \partial_{a_i}B^{(1)}\nonumber\\
R^{(1)}_{a_i 0,X}&=& -S^{(1)}_{a_i 0,X} = 0\nonumber\\
R^{(1)}_{a_i a_k,X}&=& -S^{(1)}_{a_i a_k,X}=\Psi^{(1)}\delta_{a_i a_k} - \partial_{a_k}\partial_{a_i}E^{(1)} -E^{(1)}_{a_i a_k}\nonumber
\end{eqnarray}
We can read directly from these expressions the transformation rules for the tetrad

\begin{eqnarray}\label{Ttetrads1}
\delta^{(1)}_Y e_0^{\mu}=\Tr\left(\delta^{(1)}_X e_0^{\mu}\right) &=& -\Tr(\Phi^{(1)})\eb_0^{\mu} -\eb_{a_i}^{\mu} \partial^{a_i}\Tr(B^{(1)})  \\
\delta^{(1)}_Y e_{a_i}^{\mu}=\Tr\left(\delta^{(1)}_X e_{a_i}^{\mu}\right) &=& \Tr(\Psi^{(1)})\eb_{a_i}^{\mu} - \eb_{a_k}^{\mu}\partial^{a_k}\partial_{a_i}\Tr(E^{(1)}).  \nonumber
\end{eqnarray}

\subsubsection{Second order}

\begin{eqnarray}\label{B3}
R^{(2)}_{00,X} &=& \Phi^{(2)} - 3 \Phi^2 + \dbi B \dhi B\\
R^{(2)}_{0 a_i,X} &=& - \partial_{a_i}B^{(2)} + (2 \Phi -4 \Psi)\partial_{a_i}B  + 4 \partial^{a_j}B \left(\partial_{a_i}\partial_{a_j}E + E_{a_i a_j}\right)\nonumber\\
R^{(2)}_{a_i 0,X}&=& -S^{(2)}_{a_i 0,X} = 0\nonumber\\
R^{(2)}_{a_i a_k,X}&=& -S^{(2)}_{a_i a_k,X} \nonumber\\
&=& \Psi^{(2)}\delta_{a_i a_k} - \left(\partial_{a_k}\partial_{a_i}E^{(2)} +
  E^{(2)}_{a_k a_i}\right)  + 3 \Psi^2 \delta_{a_i a_k}\nonumber\\
&& + 3 \left(\partial_{a_i}\partial^{a_l}E + E^{a_l}_{a_i}\right)
\left(\partial_{a_l}\partial_{a_k}E+E_{a_l a_k}\right) \nonumber\\
&& -6 \Psi \left(\partial_{a_i}\partial_{a_k}E+E_{a_i a_k}\right)\nonumber\\
-S^{(2)}_{00,X} &=& \Phi^{(2)} - \Phi^2 + \dbi B \dhi B\nonumber\\
-S^{(2)}_{0 a_i,X} &=& - \partial_{a_i}B^{(2)} -2 \Psi\partial_{a_i}B  + 2 \partial^{a_j}B \left(\partial_{a_i}\partial_{a_j}E+E_{a_i a_j}\right)\nonumber
\end{eqnarray}
In these formulas, we have omitted the first order superscript as there is
no possible confusion. In the following, we will also omit the first order
superscript. The transformations rules for the tetrads can be read, as we did for the first order case:

\begin{eqnarray}\label{Ttetrads2}
\Tr\left(\delta^{(2)}_X e_0^{\mu}\right) &=& -\left[\Tr(\Phi^{(2)})- 3 \Tr(\Phi)^2 + \dbi \Tr(B) \dhi \Tr(B) \right]\eb_0^{\mu} \\
&& +\Big\{ -\partial^{a_i}\Tr(B^{(2)}) + \left[2 \Tr(\Phi) -4 \Tr(\Psi)\right]\partial^{a_i} \Tr(B)  \nonumber\\
&& \qquad + 4 \partial^{a_j}\Tr(B) \left[\partial^{a_i}\partial_{a_j}\Tr(E)+E^{a_i}_{\,\,a_j}\right]\Big\}\eb_{a_i}^{\mu}   \nonumber\\
\Tr\left(\delta^{(2)}_X e_{a_i}^{\mu}\right) &=& \left[\Tr(\Psi^{(2)}) + 3 \Tr(\Psi)^2 \right]\eb_{a_i}^{\mu} \nonumber\\
&&+\Big\{-\partial^{a_k}\partial_{a_i}\Tr(E^{(2)}) +3 \left[\partial_{a_i}\partial^{a_j}\Tr(E) +E^{a_j}_{\,a_i}\right] \left[\partial^{a_k}\partial_{a_j}\Tr(E)+ E^{a_k}_{\,a_j}\right]\nonumber\\
&& \qquad - 6 \Tr(\Psi)\left[\partial^{a_k}\partial_{a_i}\Tr(E)+E^{a_k}_{\,a_i}\right]  \Big\}\eb_{a_k}^{\mu}.  \nonumber
\end{eqnarray}

\section{Transformation of $\delta^{(2)} f$}\label{app_Tf2}

\begin{eqnarray}
&&\Tr\left( \delta^{(2)}_X f\right) =\\
&&\Big\{  \left(\Lie_{T \xi_2}+\Lie^2_{T
    \xi_1}\right)\left[\bar{f}(x^{\nu},a p^{\mu}) \right] +2 \Lie_{T \xi_1}\left[\delta_X^{(1)} f(x^{\nu},a p^{\mu}) \right]  \nonumber\\
&& \quad+ \left[\Tr\left(R^{(2)c}_{a,X}\right)+S^{(2)c}_{a,X}+ 2 S^{(1)d}_{a,X}\Tr\left(R^{(1)c}_{d,X}\right)\right]\pi^a \dsd{\bar{f}}{ \pi^c} \nonumber\\
&& \quad +   \left[\Tr\left(R^{(1)b}_{a,X}\right) \Tr\left(R^{(1)d}_{c,X}\right) +S^{(1)b}_{a,X} S^{(1)d}_{c,X}+ 2 S^{(1)b}_{a,X} \Tr\left(R^{(1)d}_{c,X}\right)\right]\pi^a \pi^c   \frac{\partial^2 \bar{f}}{\partial \pi^b \partial \pi^d}  \nonumber\\
&& \quad + 2\Tr\left(R^{(1)b}_{a,X}\right)\pi^a \frac{\partial}{\partial
  \pi^b} \Lie_{T \xi_1}\left[\bar{f}(x^{\nu},a p^{\mu}) \right] + 2
\Lie_{T \xi_1} \left[S^{(1)b}_{a,X}\pi^a \frac{\partial}{\partial \pi^b} \bar{f}(x^{\nu},a p^{\mu}) \right]\nonumber\\
&&\quad + 2 \left[\Tr\left(R^{(1)b}_{a,X}\right)+ S^{(1)b}_{a,X}\right]\pi^a \frac{\partial}{\partial \pi^b} \delta_X^{(1)} f(x^{\nu},a p^{\mu}) \Big\}.\nonumber
\end{eqnarray}

These individual terms are explicitly given by

\begin{eqnarray}
&&\left[\Tr\left(R^{(2)0}_{0,X}\right)+S^{(2)0}_{0,X}+ 2
  S^{(1)0}_{0,X}\Tr\left(R^{(1)0}_{0,X}\right)\right]\pi^0 \dsd{\bar{f}}{
  \pi^0} =\\
&&\qquad \Big[-\left(T^{(2)'} + \HH T^{(2)} + S_{\Phi}(T,L)\right) + 4 \Phi (T' + \HH
  T)  + 3(T' + \HH T)^2 \nonumber\\
&& \qquad - 2 \dbi B \dhi(-T+L') - \dbi (-T+L')\dhi (-T+L') \Big] \pi^0 \dsd{\bar{f}}{\pi^0}, \nonumber
\end{eqnarray}

\begin{eqnarray}
 &&\left[\Tr\left(R^{(1)0}_{0,X}\right) \Tr\left(R^{(1)0}_{0,X}\right)
   +S^{(1)0}_{0,X} S^{(1)0}_{0,X}+ 2 S^{(1)0}_{0,X}
   \Tr\left(R^{(1)0}_{0,X}\right)\right]\pi^0 \pi^0   \frac{\partial^2
   \bar{f}}{\partial \pi^0 \partial \pi^0}  =\nonumber\\
&& \qquad \qquad \frac{\partial^2 \bar{f}}{\partial \left(\pi^0 \right)^2}(\pi^0)^2\left(T' + \HH T\right)^2,
\end{eqnarray}

\begin{eqnarray}
&&2 \left[\Tr\left(R^{(1)b}_{a,X}\right)+ S^{(1)b}_{a,X}\right]\pi^a
\frac{\partial}{\partial \pi^b} \delta_X^{(1)} f(x^{\nu},a p^{\mu}) =\\
&& \qquad  -2 \dsd{\delta_X^{(1)} f}{\pi^0} \pi^0 (T' + \HH T)- 2  \dsd{\delta_X^{(1)} f}{\pi^i} \pi^0 (- \dhi T + \dhi L') \nonumber\\
&& \qquad- 2  \dsd{\delta_X^{(1)} f}{\pi^i} (\pi^j  \dhi \dbj L + \HH \pi^i T ), \nonumber
\end{eqnarray}

\begin{eqnarray}
&&2\Tr\left(R^{(1)b}_{a,X}\right)\pi^a \frac{\partial}{\partial \pi^b}
\Lie_{T \xi_1} \left[\bar{f}(x^{\nu},a p^{\mu}) \right] =\\
&& \qquad - 2 \left[ \frac{\partial^2
    \bar{f}}{\partial \left(\pi^0 \right)^2}(\pi^0)^2  + \dsd{\bar{f}}{\pi^0}
  \pi^0\right]\left(T' + \HH T\right)(\Phi+ T' + \HH T)\nonumber\\
&& \qquad - 2 \frac{\partial^2 \bar{f}}{\partial \left(\pi^0 \right)^2} (\pi^0)^2 \left(n^i \dbi T
\right)(\Phi+ T' + \HH T)- 2 T \frac{\partial^2 \bar{f}}{\partial \eta
  \partial \pi^0}(\Phi + T'+ \HH T)\nonumber\\
&& \qquad - 2  \dsd{\bar{f}}{\pi^0} \pi^0 (B^i - \dhi T + \dhi L') \dbi T  \nonumber\\
&& \qquad- 2  \dsd{\bar{f}}{\pi^0} \pi^0 \left[n^j \dhi
  \dbj(E+L)+n^j E^i_{\,\,j} + n^i(-\Psi+\HH T)\right] \dbi T,  \nonumber 
\end{eqnarray}

\begin{eqnarray}
&&2 \Lie_{T \xi_1}\left[S^{(1)b}_{c,X}a \pi^c \frac{\partial}{\partial
    \pi^b} \bar{f}(x^{\nu},a p^{\mu}) \right]= \\
&& \qquad 2\frac{\partial^2 \bar{f}}{\partial \eta  \partial \pi^0} \pi^0 \Phi T + 2\dsd{\bar{f}}{\pi^0}\pi^0 \left(\Phi' T + \dbi \Phi \dhi L\right)\nonumber\\
&& \qquad + 2\left[ \frac{\partial^2 \bar{f}}{\partial \left(\pi^0 \right)^2}(\pi^0)^2  + \dsd{\bar{f}}{\pi^0} \pi^0\right] \Phi \left(T' + \HH T + n^i \dbi T\right),\nonumber
\end{eqnarray}

\begin{eqnarray}
 &&\left(\Lie_{T \xi_2}+\Lie^2_{T
    \xi_1}\right)\left[\bar{f}(x^{\nu},a p^{\mu}) \right]= \\
&&\quad T^{(2)} \frac{\partial \bar{f}}{\partial \eta} + \frac{\partial \bar{f}}{\partial \pi^0}\pi^0
(T^{(2)'}+\HH T^{(2)}+ n^i \dbi T^{(2)})\nonumber\\
&&\quad  + \frac{\partial^2 \bar{f}}{\partial
  \eta^2}T^2 +\dsd{\bar{f}}{\eta}(T T'+ \dbi T \dhi L) + \frac{\partial^2
  \bar{f}}{\partial \pi^0 \partial \eta} \pi^0 2T (T'+\HH T +n^i \dbi T)\nonumber\\
&& \quad +\frac{\partial^2 \bar{f}}{\partial \left(\pi^0 \right)^2}(\pi^0)^2\left[ 2 n^i \dbi T (\HH T + T') + (n^i \dbi T)(n^j \dbj T)
  +\HH^2 T^2 + 2 \HH T T'+ (T')^2 \right]  \nonumber\\
&& \quad + \dsd{\bar{f}}{\pi^0} \pi^0 \Big[T T'' + \HH'T^2 + 3\HH T T'+ T n^i \dbi T'+
  T' n^i \dbi T + \HH^2 T^2 + 2 \HH T n^i \dbi T \nonumber\\
&& \qquad \qquad + \dbj T' \dhj L + \dbj T \dhj L'+ \HH \dbj T \dhj L + \dhj L n^i \dbi \dbj T + \dhj T n^i \dbi \dbj L + (T')^2 \Big],\nonumber
\end{eqnarray}
\begin{eqnarray}
&&2 \Lie_{T \xi_1}\left[\delta_X^{(1)} f(x^{\nu},a p^{\mu})\right] =\\
&& \qquad  2 \left( \pi^0 T' + \pi^j \dbj T\right)
\dsd{\delta_X^{(1)} f}{\pi^0}  +2 \left( \pi^0 \dhi L'+ \pi^j \dhi\dbj L \right) \dsd{\delta_X^{(1)} f}{\pi^i}  \nonumber \\
&& \qquad + 2\dhi L \dsd{\delta_X^{(1)} f}{x^i} + 2T \left(\dsd{\delta_X^{(1)} f}{\eta}+\dsd{\delta_X^{(1)} f}{\pi^0}\HH \pi^0+\dsd{\delta_X^{(1)} f}{\pi^i}\HH \pi^i\right). \nonumber
\end{eqnarray}

In the above formulas, we have omitted to write the fact that the derivatives with respect to $\eta$ or $x^i$ are taken at fixed $\pi^a$.

\section{Integral relations necessary to derive the fluid limit}\label{app_int_utiles}

The integrations on angular directions can be handled with the general formulas
(see Ref.~\cite{Uzan98})

\begin{equation}
\int n^{i_1}...n^{i_k}\frac{\dd^2 \Omega}{4 \pi}=0 \quad {\rm if} \quad k= 2p + 1 
\end{equation}

\begin{equation}
\int n^{i_1}...n^{i_k}\frac{\dd^2 \Omega}{4 \pi}=\frac{1}{k+1}\left(\delta^{(i_1
    i_2}...\delta^{i_{(k-1)}i_k)}\right) \quad {\rm if} \quad k= 2p. 
\end{equation}

By successive integration by parts, we also obtain the following useful results

\begin{eqnarray}
\int \bar{f}(x^{\mu},\pi^0) (\pi^0)^3 \dd \pi^0 \dd^2 \Omega &=& \bar{\rho}(x^{\mu}),\nonumber\\
\int \dsd{\bar{f}(x^{\mu},\pi^0)}{\pi^0}(\pi^0)^4 \dd \pi^0 \dd^2 \Omega  &=& -4\bar{\rho}(x^{\mu}),\nonumber\\
\int \frac{\partial^2 \bar{f}(x^{\mu},\pi^0)}{\partial^2 \pi^0}(\pi^0)^5 \dd \pi^0 \dd^2 \Omega &=& 20 \bar{\rho}(x^{\mu}).
\end{eqnarray}

\section{The fluid limit for radiation}\label{app_fluid_equations}

As explained in section \ref{sec1part1}, second order quantities involve
either purely second order perturbation variables or terms quadratic in first
order perturbation variables. As long as the order of the quantity is known we
can omit the order superscript in order to simplify notations. 

\subsection{Geometric quantities}

In the Newtonian gauge, ignoring vector perturbations for simplicity, the non-vanishing Christoffel symbols associated with the metric (\ref{metric}) are for the background

\begin{equation}\label{eq:A1}
 \ChristoffelOrdre{0}{0}{0}{(0)} = \HH,\,\,
 \ChristoffelOrdre{0}{j}{k}{(0)} = \HH \delta_{jk},\,\,
 \ChristoffelOrdre{i}{0}{j}{(0)}= \HH \delta^{i}_{j}\,.
\end{equation}
At first order, we get
\begin{equation}
 \ChristoffelOrdre{0}{0}{0}{(1)} = \Phi',\,\,\,
  \ChristoffelOrdre{0}{0}{j}{(1)} = \partial_{j}\Phi,\,\,\,
 \ChristoffelOrdre{i}{0}{0}{(1)} = \dhi \Phi,
\end{equation}
\begin{eqnarray}
 \ChristoffelOrdre{0}{j}{k}{(1)} &=&
 2 \HH E_{jk}+ E_{jk}' -\left( 2 \HH \Phi +\Psi' + 2 \HH\Psi\right) \delta_{jk},  \\
 \ChristoffelOrdre{i}{0}{j}{(1)} &=& {E'^i}_{j}-\Psi' \delta^i_j,
 \end{eqnarray}
\begin{equation}
 \ChristoffelOrdre{i}{j}{k}{(1)} =
 2 \dv{(k}[E^i_{\,j)}- \Psi \delta^i_{j)}] - \partial^{i}(E_{jk}-\Psi
\delta_{jk}),
\end{equation}
where $A_{(ij)}\equiv (A_{ij}+A_{ji})/2$.
At second order, we obtain
\begin{equation}
 \ChristoffelOrdre{0}{0}{0}{(2)} = \Phi' - 4\,\Phi\,\Phi',\,\,
 \ChristoffelOrdre{0}{0}{j}{(2)} = \partial_{j}\Phi-
 4 \,\Phi\,\partial_{j}\Phi,
\end{equation}
\begin{equation}
 \ChristoffelOrdre{i}{0}{0}{(2)} = \dhi \Phi - 4 E^{ij} \dbj \Phi + 4 \Psi \dhi \Phi,
\end{equation}
\begin{eqnarray}
 \ChristoffelOrdre{0}{j}{k}{(2)} &=&
  \left[- 2 \HH \Psi - \Psi' + 4 \Phi \Psi' -2 \HH\Phi + 8 \HH \Phi \left(\Phi + \Psi\right) \right]\delta_{jk}\nonumber\\
 &&  + 2\HH E_{jk} - 8 \Phi\HH E_{jk} + E_{jk}' - 4 \Phi E_{jk}',
\end{eqnarray}
\begin{eqnarray}
 \ChristoffelOrdre{i}{0}{j}{(2)} &=& E^{i'}_{\,j}+ 4 \Psi' E^{i}_{\,j}
 - \Psi' \delta^i_j - 4 \Psi \Psi' \delta^i_j\nonumber\\
 &&- 4 E^{ik}E'_{kj} + 4 \Psi {E^{i}_{\,j}}',
\end{eqnarray}
\begin{eqnarray}
 \ChristoffelOrdre{i}{j}{k}{(2)} &=&  2 \dv{(k}[E^i_{\,j)}- \Psi \delta^i_{j)}] - \partial^{i}(E_{jk}-\Psi \delta_{jk})\\
&& + 4 \big(E^{il}-\Psi \delta^{il}\big) \Big[  \dv{l}( E_{jk}-\Psi
  \delta_{jk} )   \nonumber\\
&& \qquad \qquad \qquad - \dv{k}( E_{lj}-\Psi \delta_{lj} ) -\dv{j}( E_{kl}-\Psi \delta_{kl} )   \Big].\nonumber
\end{eqnarray}
 \subsection{The radiation fluid equations}

The conservation equation $\nabla_{\mu} T^{\mu \nu}$ for the stress
energy tensor (\ref{defstressenergytensor}) with a radiation
equation of state $P=\rho/3$ and where we assume $\pi^{\mu
  \nu} =0$, are the conservation equation 
\begin{eqnarray}\label{Eqcons1}
\left(\delta^{(1)} \rho\right)'+ 4 \HH \delta^{(1)} \rho + \frac{4}{3}\bar{\rho} \left(\Delta v^{(1)} -3 \Psi^{(1)\prime}\right) & = & 0,\\
 \left(\delta^{(2)} \rho\right)'+ 4 \HH \delta^{(2)} \rho + \frac{4}{3}\bar{\rho} \left(\Delta v^{(2)} -3 \Psi^{(2)\prime}\right)& = & S_{c},\nonumber
\end{eqnarray}
and the Euler equation 
\begin{eqnarray}\label{Eqcons2}
 v^{(1)\prime}+ \Phi^{(1)} + \frac{\delta^{(1)} \rho}{4 \bar{\rho}} & = & 0,\\
 v^{(2)\prime}+ \Phi^{(2)} + \frac{\delta^{(2)} \rho}{4 \bar{\rho}} & = & S_{e},\nonumber
\end{eqnarray}
where the source terms in the second order equations, which are quadratic in
first order perturbation variables, are given by
 \begin{eqnarray}
 S_{c} &=& \frac{8}{3} \Big\{ \delta \rho \Psi' + 6 \bar{\rho}\Psi \Psi' -
   (\Phi+\delta)\bar{\rho}\Delta v \nonumber\\
&& \qquad + \partial_i v\left[-\partial^i\delta \rho -2 \bar{\rho}\partial^i v'-2\bar{\rho}\partial^i\Phi+3\bar{\rho}\partial^i\Psi\right]\Big\},\\
 \dbi S_{e} &=& -2 \left(\frac{\delta \rho}{\bar{\rho}} \dbi v\right)' + 10 \Psi' \dbi v
               + 4 \Psi \dbi v'-2 \dbj \big( \dhj v \dbi v \big) \nonumber\\
&& \qquad  + 2 \Phi \dbi v'-2 \frac{\delta \rho}{\rho} \dbi \Phi + 4 \Phi \dbi \Phi.
\end{eqnarray}


\end{document}